\documentclass[%
 reprint,
  superscriptaddress,
 amsmath,amssymb,amsfonts,
 aps,
 prl,
 floatfix,
 longbibliography
]{revtex4-2}

\usepackage{bm}

\usepackage[pdftex]{graphicx} \graphicspath{{}}
\usepackage{float} \usepackage{color}

\usepackage[pdftex,colorlinks=true]{hyperref}
\hypersetup{
  colorlinks=true,
  linkcolor=blue,
  urlcolor=cyan,
}

\usepackage{mathtools}

\DeclarePairedDelimiterX\braket[2]{\langle}{\rangle}{#1 \delimsize\vert #2}
\DeclarePairedDelimiterX\expval[3]{\langle}{\rangle}{#1 \delimsize\vert #2  \delimsize\vert #3}
\DeclarePairedDelimiterX\proj[2]{\delimsize\vert#1\rangle}{\langle#2\delimsize\vert}{ }

\usepackage{siunitx}

\newcommand{\vect}[1]{\mathbf{#1}}

\begin{document}

\title{Manipulating growth and propagation of correlations in dipolar multilayers: 
From  pair production to bosonic Kitaev models}

\author{Thomas Bilitewski}
\affiliation{Department of Physics, Oklahoma State University, Stillwater, Oklahoma 74078, USA}
\author{Ana Maria Rey}
\affiliation{JILA, National Institute of Standards and Technology and Department of Physics, University of Colorado, Boulder, CO, 80309, USA}
\affiliation{Center for Theory of Quantum Matter, University of Colorado, Boulder, CO, 80309, USA}

\date{\today}

\begin{abstract}
We study the non-equilibrium dynamics of dipoles confined in multiple stacked two-dimensional layers realising a long-range interacting quantum spin 1/2 XXZ model. We demonstrate that strong in-plane XXX interactions can protect a manifold of collective layer dynamics. This then allows us to map the many-body spin dynamics to bosonic models. In a bilayer configuration we show how to engineer the paradigmatic two-mode squeezing Hamiltonian known from quantum optics, resulting in exponential production of entangled pairs and generation of metrologically useful entanglement from initially prepared product states. In multi-layer configurations we engineer a bosonic variant of the Kitaev model displaying chiral propagation along the layer direction. %
Our study illustrates how the control over interactions, lattice geometry and state preparation in interacting dipolar systems uniquely afforded by AMO platforms such as  Rydberg and magnetic atoms, polar molecules or trapped ions  allow for the control over  the temporal and spatial propagation  of correlations for  applications in quantum sensing and quantum simulation.
\end{abstract}

\maketitle


The individual particle control  recently offered by quantum gas microscopes \cite{NatPhysGross2021} and optical tweezers \cite{NatPhysKaufman2021}, the impressive  advances in spectroscopic methods \cite{NatPhysVale2021}, complemented by the  capability of experiments to trap  and manipulate a broad range of atomic, molecular and optical systems  featuring  diverse  types of interactions ( from contact  \cite{NPhysBloch2005,RevModPhys.80.885}, to dipolar   \cite{Bohn_Science_357_2017,Moses2016,Baranov_ChemicalReviews_112_2012,NatPhysBrowaeys2020,QScMorgado2021, RepProgPhysLahaye2009,Chomaz2022,ArxivDefenu2021} to all-to-all \cite{AiPMivehvar2021,ScienceNorcia2018,PRLParkings2017,PRLSchleierSmith2019,NatPhysBlatt2012,RevModMonroe2021}) are opening untapped  opportunities for quantum simulation \cite{NPhysBloch2012,ScienceGross2017,NatureDaley2022}, metrology \cite{RevModPhys.90.035005} and computation \cite{Briegel_2000,Weiss2017,Henriet2020quantumcomputing}. 
In  these systems it is now possible to explore the    propagation and growth of quantum entanglement and correlations \cite{NatRevPhys_Lewis_Swan_2019} which is crucial for demonstrating their quantum advantage. 

One of  the most basic mechanisms for entanglement growth, which is also at the very heart of foundational questions in quantum mechanics \cite{EPR_1935,RevModPhys.81.1727}, is the creation of entangled states consisting of pairs of correlated particles in the guise of two-mode squeezed (TMS) states \cite{agarwal2013quantum,Caves1985,Schumaker1985}. These states were originally understood in quantum optics in the context of parametric amplification, but  have been shown to be relevant to  a wide range of phenomena including  the Schwinger effect in  high energy physics \cite{Hauke_PhysRevX_2013,Kasper_PhysLettB_2016}, the Unruh thermal radiation in general relativity \cite{Hu_NatPhys_2019}, mode-changing collisions in spinor condensates \cite{Gross_Nature_2011,Lcke_Science_2011,Linneman_PRL_2016,Qu_PRL_2020} and thermofield double states  in the holographic correspondence relating a quantum-field theory to a gravitational theory in one higher dimension \cite{Chapman_SciPost_2019,Zhu_PNAS_2020}.

\begin{figure}
\includegraphics[width=\columnwidth]{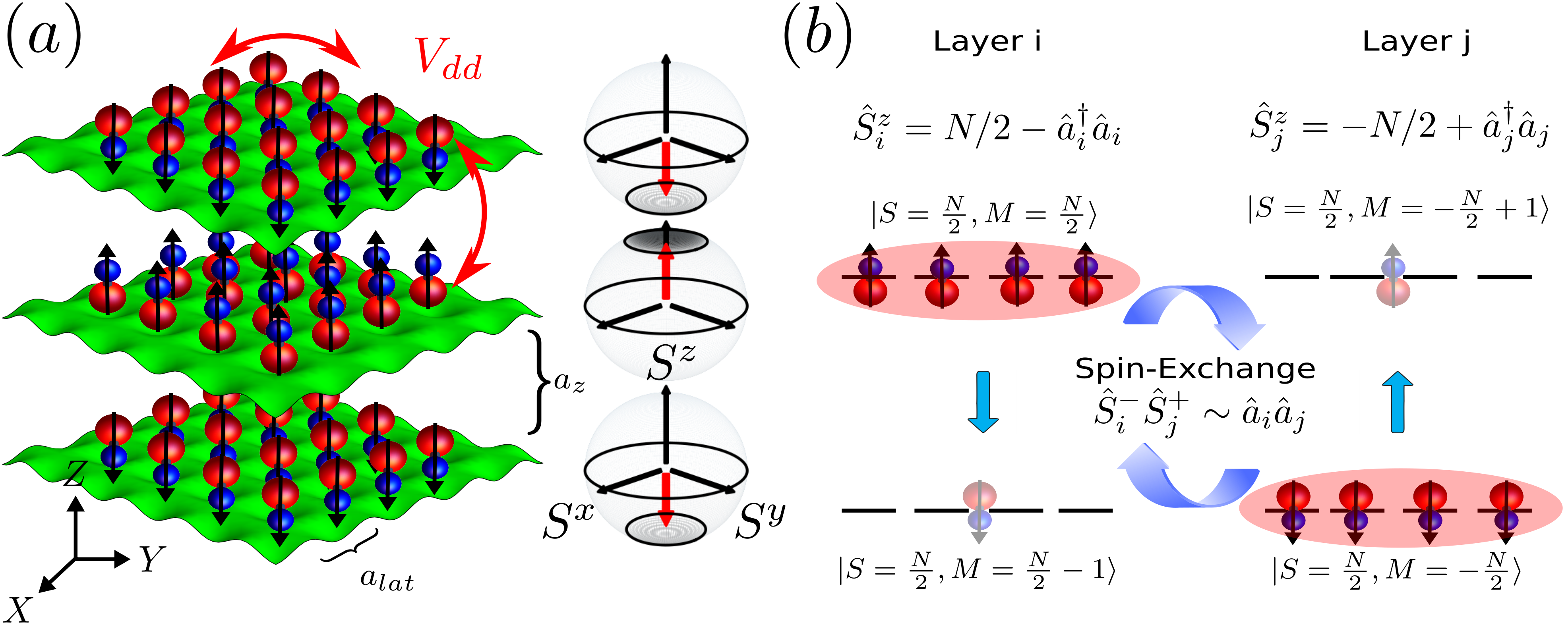}
\caption{Illustration of spin 1/2 dipoles in multi-layers and mapping to bosonic pair creation. a) Dipoles in stacked 2D layers of a 3D optical lattice are prepared layer selectively in (distinct) coherent spin states illustrated by the Bloch spheres. They interact via long-range (dipolar) interactions within and between layers.  b) Strong in-plane interactions $\hat{S}_i \cdot \hat{S}_i$ (illustrated as red ellipse) couple spins within a layer resulting in collective behavior. Inter-layer dipolar spin exchange ($\hat{S}_i^+ \hat{S}_j^-$) maps to pair creation of bosonic collective excitations.\label{fig:model}}
\end{figure} 

In this work we explore various ways to produce correlated pairs during during  the non-equilibrium many-body dynamics  of spins 1/2 arrays  prepared in a stack of   two-dimensional  layers. Specifically, we use strong in-plane Heisenberg interactions to lock the spin of each layer into a collective spin, with magnitude set by the number of particles in each layer, and show that by  preparing different initial orientations of the collective spins (enabled by layer-selectivity \cite{Tobias_Science_375_2022}), interlayer interactions \cite{Tobias_Science_375_2022} can be used to engineer distinct types of pair production processes.

One of them is the paradigmatic  two-mode squeezing Hamiltonian \cite{agarwal2013quantum,Schumaker1985,Caves1985} known for its capability to generate   metrologically useful states. %
 Originally encapsulated by the Einstein, Podolsky, and Rosen (EPR) paradox \cite{EPR_1935,RevModPhys.81.1727}, two-mode squeezing occurs when two separate ensembles A and B are correlated such that the relative fluctuations between the  sum and difference of two quadratures  can be determined below the Heisenberg uncertainty constraint \cite{Polzik_RevModPhys_2010}. In this work A and B are bosonic excitations in different spatially separated layers generated exponentially fast through long-range dipolar interactions between the particles allowing fast scalable entanglement generation in large systems. %
%

Pioneering work on spatially distributed entanglement has been accomplished using atom-light interactions in photonic systems \cite{freedman1972experimental,aspect1981experimental,ou1992realization}, and hot atomic vapors  \cite{julsgaard2001experimental,cerf2007quantum}, as well as  in Bose-Einstein condensates \cite{Fadel2017,Kunkel2018}, and more generally in quantum networks \cite{PhysRevLett.124.110501,Nichol_2022,Wan_2019,Lago_Rivera_2021}. However, the possibility of using long-range interactions to directly correlate spatially separated arrays without the  detrimental degradation of coherence from motional effects or photon loss can offer significant opportunities for quantum metrology, in particular in terms of scalability and speed of entanglement generation. Furthermore, by preparing an initial state with states cyclically staggered along three perpendicular Bloch vector directions we show it is possible to engineer a bosonic variety \cite{Zhou_2011,McDonald_PhysicalReviewX_8_2018} of the Kitaev  model \cite{Kitaev_2001} which shows remarkable properties such as phase-dependent chirality, drastic sensitivity to boundary conditions and rich dynamical behaviors \cite{McDonald_PhysicalReviewX_8_2018}.
While there have been proposals to generate  bosonic Kitaev models in coupled cavities subject to parametric driving \cite{McDonald_PhysicalReviewX_8_2018}, their implementation in long lived molecular or atomic  states  interacting via long range  interactions  can offer important advantages for their preparation, detection and storage .

{\it Model.---}
We consider spins interacting via long-range interactions in two or more two-dimensional layers as shown in Fig.~\ref{fig:model}(a), prepared for example via a  deep 3D optical lattice .  We assume distinct in-plane lattice spacing $a_{\mathrm{lat}}$ and layer spacing $a_Z$. We restrict dynamics to two internal states representing a spin 1/2 degree of freedom with  dynamics determined by the XXZ Hamiltonian
\begin{equation}
    \hat H_{\mathrm{XXZ}} =1/2 \sum_{\mathbf{i}\neq \mathbf{j}} V_{\mathbf{i}\mathbf{j}}\left[\frac{J_{\perp}}{2} (\hat s_{\mathbf{i}}^+ \hat s_{\mathbf{j}}^- + \hat s_{\mathbf{i}}^- \hat s_{\mathbf{j}}^+) + J_z \hat s_{\mathbf{i}}^z \hat s_{\mathbf{j}}^z \right] 
\end{equation}
where $\mathbf{i},\mathbf{j}$ are three-dimensional  positions $(i_X,i_Y,i_Z)$ and $i_X,i_Y$ run along the positions in a given two-dimensional layer of size $L\times L$ indexed by $i_Z$. The spin-operators $\hat s_{\mathbf{i}}^\alpha = \hat \sigma_{\mathbf{i}}^\alpha / 2$ are given in terms of the Pauli matrices $\hat \sigma^{x,y,z}$ that act on the spin  at site $\mathbf{i}$. The couplings $J_{\perp}$ and $J_z$ determine the relative strength of the spin-exchange and Ising terms respectively, which we will tune to be equal, $J_z=J_{\perp}$ in the following. 

For specificity we consider dipolar interactions of the form $V_{\mathbf{i}\mathbf{j}} =  V_{dd}(\vect{r}_{\mathbf{i}} - \vect{r}_{\mathbf{j}})$ with $V_{dd}(\vect{r}) =  \frac{C_{dd}}{r^3} (1 -3 Z^2)$ parametrised by a dipolar coupling strength $C_{dd}$. We note that the Hamiltonian contains both intra-layer as well as inter-layer Heisenberg interactions, the relative strength of which we can tune by changing the ratio $a_Z/a_{\mathrm{lat}}$. We emphasise that the specific spatial dependence of the interactions is not material to our conclusions as long as intra-plane versus inter-plane interactions are tunable, and the interactions protect collective behaviour which can already occur for nearest neighbour interactions \cite{Perlin_PhysicalReviewLetters_125_2020}.

We assume that in a given layer $i_Z$ all spins are prepared in the same state such that the collective layer spin $\vec{S}_{i_Z}^2 = \sum_{i_X,i_Y} \vec{s}_{i_X,i_Y,i_Z}$  points in a layer dependent direction $\langle\vec{S}_{i_Z} \rangle = N/2 \, \vec{n}_{i_Z}$ with $N=L^2$ at unit filling as illustrated in Fig.~\ref{fig:model}(a) for alternating anti-aligned layer spins, $\vec{n}_{i_Z}= (-1)^{i_Z} \hat{z}$. For simplicity we have assumed all the layers have the same particle number $N$.

At the Heisenberg point, achieved when $J_z=J_{\perp}$, the interactions between spins within a plane open a many-body gap between permutationally symmetric states of maximal spin-length $\langle \vec{S}_{i_Z}^2\rangle=N/2 (N/2+1)$ and states with $\langle\vec{S}_{i_Z}^2\rangle  < N/2 (N/2+1)$ \cite{Rey_Phys.Rev.A_77_2008,Cappellaro2009,Kwasigroch2014,Davis2020,Perlin_PhysicalReviewLetters_125_2020}. Assuming the dynamics remains in this manifold the model further reduces to 
\begin{equation}
    \hat H_{layer} =1/2 \sum_{i \neq j} V^{\mathrm{av}}_{ij}\left[\frac{1}{2} (\hat S_i^+ \hat S_j^- + h.c.) + \hat S_i^z \hat S_j^z \right] 
\end{equation}
where we have removed for simplicity the subscript $z$, $i,j$ now denote  the layers, and $V^{\mathrm{av}}_{ij} = 1/N^2 \sum_{x,y,x',y'} V_{(x,y,i),(x',y',j)}$. Note that we dropped the in-plane interactions, which in the fully collective manifold just contribute a constant term. In this collective limit the layer-averaged interaction also sets the natural scale for the time evolution. For convenience we define the nearest-layer averaged interaction $V = V_{i,i+1}^{\mathrm{av}}$.

{\it Bilayer.---}
\begin{figure}
\includegraphics[width=\columnwidth]{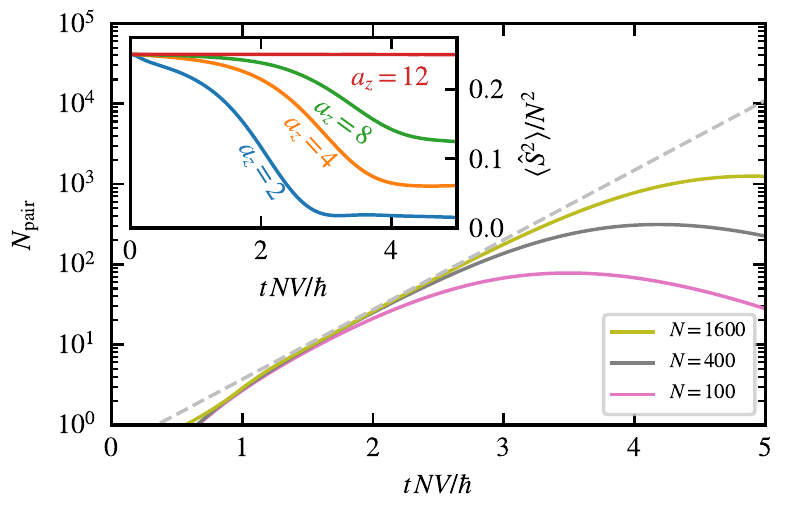}
\caption{Exponential pair-creation and dynamical phase transition in a dipolar bilayer. Main panel: Exponential growth of bosonic excitations, $N_{\mathrm{\mathrm{pair}}} = S^z_1-S^z_0-N$, versus time compared to the two-mode squeezing prediction $N_{\mathrm{pair}} = 2\sinh^2{( S Vt/\hbar)}$ (gray dashed). Bilayer with $a_Z/a_{\mathrm{lat}}=12$ in an initially anti-aligned spin state, $S_1^z(0) = - S_0^z(0) = N/2$ with $L=10,20,40$ ($N=100,400,1600$ per layer). Inset: DPT to collective regime. Spin-length $\langle \hat{S}^2\rangle/N^2$ versus time for different layer spacings $a_Z/a_{\mathrm{lat}}=2,4,8,12$ displaying the transition to collective behavior for $L=40$ ($N=1600$).
\label{fig:pair_creation}}
\end{figure} 
We first study the case of a bilayer configuration with initially anti-aligned layer spins, $\langle\vec{S}_{1}(t=0) \rangle = 
-\langle\vec{S}_{0}(t=0) \rangle=N/2 \, \hat{z}$, as shown in Fig.~\ref{fig:model}(b).

We simulate the quantum dynamics of the full dipolar spin model using the discrete truncated Wigner approximation (dTWA) \cite{Schachenmayer_Phys.Rev.X_5_2015,Zhu_NewJournalofPhysics_21_2019,Sundar2019,Wurtz2018,Kolodrubetz2017,Berg2009}. 
In the inset of Fig.~\ref{fig:pair_creation} we show the time evolution of the total layer spin-length $\langle \hat{S}_i^2\rangle$ for a range of layer-spacings $a_Z/a_{\mathrm{lat}}$. While for closely spaced layers the dynamics quickly leaves the fully collective manifold resulting in rapid decay, for sufficiently large spacings we observe a transition to robust collective behaviour with the spin-length remaining maximal throughout the dynamics. This is readily explained by the relative increase of the in-plane interactions, which gap protect the permutationally symmetric manifold in each layer, compared to the inter-plane interactions, which allow excitations out of this manifold,  with increasing layer distance. 

As a first result of this work we thus observe that for an appropriate ratio of $a_Z/a_{lat}$ (or generically for sufficiently strong intra-plane and sufficiently homogeneous inter-layer interactions) the dynamics of the full model indeed closely follows the collective model, thus allowing the simulation of 1D Heisenberg chains with large (tunable) spin.

{\it Mapping to two-mode Squeezing.---}
Having demonstrated the collectiveness during the dynamics, we use a standard Holstein-Primakoff transformation \cite{Holstein1940} as $S_1^z = S - \hat{a}^{\dagger}a$, $S_1^+ = \sqrt{2S} \sqrt{1-\frac{\hat{a}^{\dagger}\hat{a}}{2S}} \hat{a}$, $S_1^{-} =(S_1^{+})^{\dagger}$, and $S_0^z = -S + \hat{b}^{\dagger}b$, $S_0^+ = \sqrt{2S} \hat{b}^{\dagger} \sqrt{1-\frac{\hat{b}^{\dagger}\hat{b}}{2S}} $, $S_0^{-} = (S_1^{+})^{\dagger}$ with $S=N/2$, with $\hat{a}$ and $\hat{b}$ bosonic operators. To quadratic order we obtain 
\begin{equation}
    H_{\mathrm{pair}} = S V \, \left( (\hat{a}^{\dagger} \hat{b}^{\dagger} + \hat{a} \hat{b}) + (\hat{a}^{\dagger} \hat{a} + \hat{b}^{\dagger} \hat{b} ) \right) 
    \label{eq:H_tms}
\end{equation}
where  $S_0^+ S_1^{-} + S_0^- S_1^{+} $ maps to pair creation $ 2S (\hat{a}^{\dagger} \hat{b}^{\dagger} + \hat{a} \hat{b})$, while $S_0^z S_1^{z}$ maps to $-S^2 + S (\hat{a}^{\dagger} \hat{a} + \hat{b}^{\dagger} \hat{b} )$. To cancel this Ising induced term we apply an additional staggered field $\sum_i (-1)^i h S_i^z$ with $h=-S V$ obtaining the pure two-mode squeezing Hamiltonian. Then, the dynamics corresponds to the resonant creation of correlated pairs of bosonic excitations in both layers. This mapping and the resulting pair creation due to dipolar interlayer spin exchange in the collective spin manifold is also illustrated in Fig.~\ref{fig:model}(b).

{\it Exponential Pair Creation}
A first prediction of this mapping is the exponential creation of bosonic pairs of excitations $N_{\mathrm{pair}} = (\hat{a}^{\dagger} \hat{a} + \hat{b}^{\dagger} \hat{b})=S^z_1-S^z_0-N$ as $N_{\mathrm{pair}}(t) =2 \sinh^2{(S Vt/\hbar)}$ \cite{agarwal2013quantum,Schumaker1985,Caves1985}. In Fig.~\ref{fig:pair_creation} we demonstrate that the dynamics of the full dipolar bilayer based on dTWA simulations also shows exponential creation of pairs. In fact, it closely follows the prediction of the two-mode squeezing Hamiltonian (dashed gray line) as long as $N_{\mathrm{pair}}\lesssim \sqrt{N}$, beyond which the Holstein-Primakoff approximation is invalid, and higher order corrections become relevant.


{\it Squeezed Quadratures.---}
\begin{figure}
\includegraphics[width=\columnwidth]{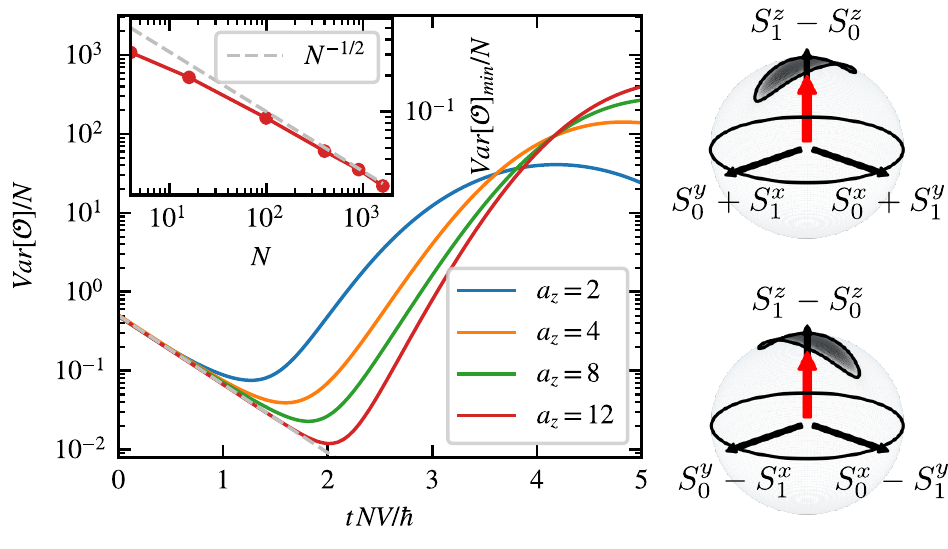}
\caption{Exponential two-mode squeezing. Main panel: Time evolution of the squeezed variances for different layer-spacings, compared to the two-mode-squeezing prediction $Var[\mathcal{O}] = N/2  e^{\pm 2 S V t /\hbar}$ (gray dashed) for $L=40$ ($N=1600$ per layer). Bloch spheres using the appropriate combinations of the layer-spin operators illustrating squeezed and anti-squeezed variances. Inset: N-dependence of the minimal variance achieved ($a_Z/a_{lat}=12$) demonstrating $N^{-1/2}$-scaling (gray dashed).
\label{fig:squeezing}
}
\end{figure} 
Due to the correlated creation of pairs in two modes the Hamiltonian (Eq.~\ref{eq:H_tms}) generates squeezed states in hybrid quadratures \cite{agarwal2013quantum,Schumaker1985,Caves1985}. Translating these well known results from the bosonic operators into our original spin operators we find that  $S_0^x + S_1^y$ and $S_0^y-S_1^x$ correspond to squeezed quadratures, and $S_0^x-S_1^y$ and $S_0^y + S_1^x$ correspond to the anti-squeezed quadratures. Consequently, the variance of these hybrid operators is predicted to evolve as $Var[\mathcal{O_{\pm}}] =N/2 e^{\pm 2 S V t /\hbar}$, where we use $\pm$ to refer to the anti-squeezed/squeezed quadratures.

Based on the full dynamics of the dipolar bilayer we confirm this prediction in Fig.~\ref{fig:squeezing} which shows the exponential decrease of the variance of the squeezed quadratures for a range of layer spacings (anti-squeezed quadratures not shown also behave accordingly). We observe that the minimal squeezing achievable relies on a sufficiently large layer separation to ensure we stay in the fully collective manifold. If that is the case we observe excellent agreement up to a time at which $N_{\mathrm{\mathrm{pair}}} \sim \sqrt{N}$ where corrections to the quadratic Hamiltonian become relevant. As a consequence, the minimal achievable squeezing scales as $N^{-1/2}$ with respect to the total number of spins as shown in the inset of Fig.~\ref{fig:squeezing}. We note that these generated states directly allow quantum-enhanced sensing using recently devised Ramsey protocols only requiring measurements of the collective spin variables and collective spin rotations of the individual layers \cite{Sundar2022}.

{\it Bosonic Kitaev Model and chiral spin transport. ---}
\begin{figure}
\includegraphics[width=\columnwidth]{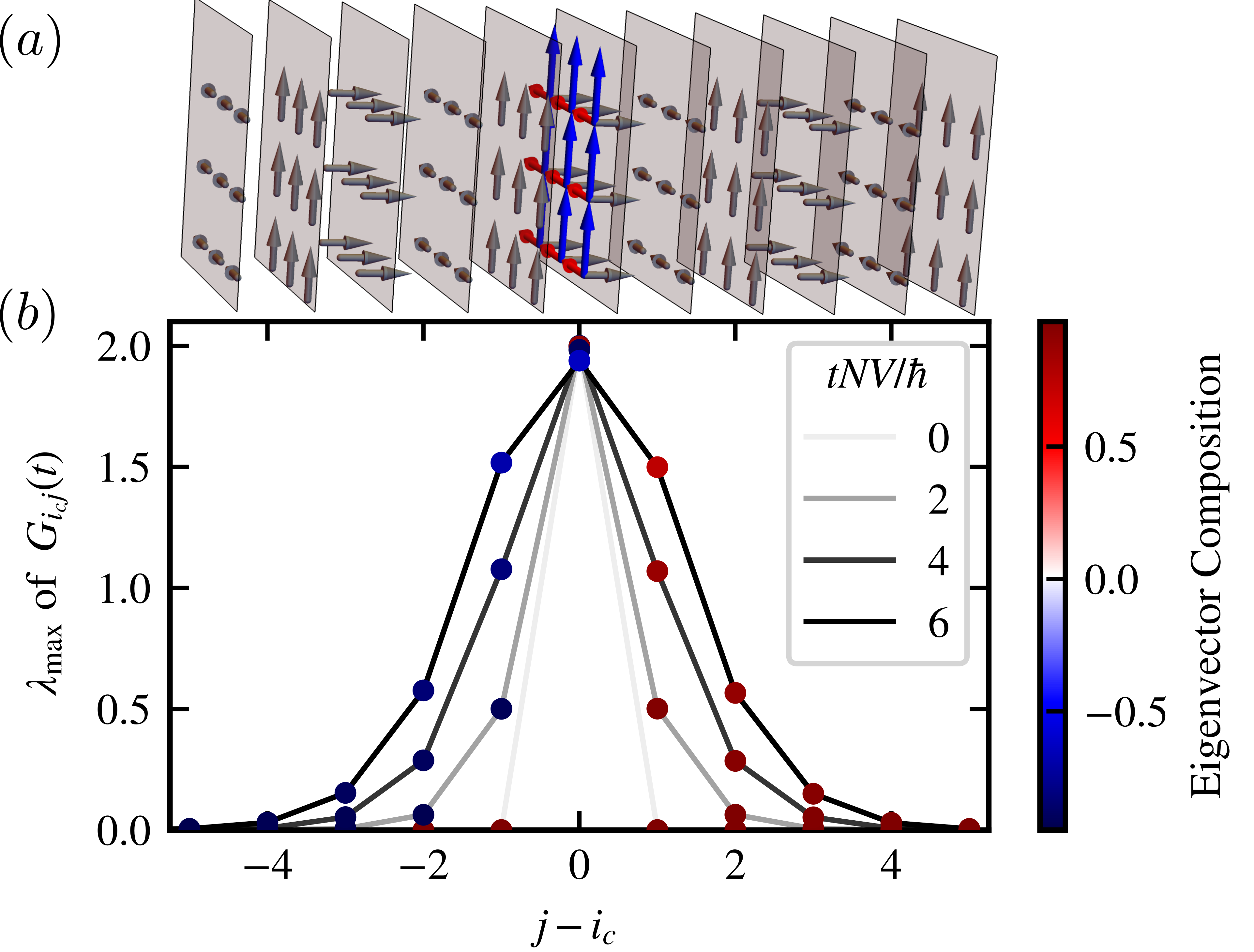}
\caption{Chirality of spin transport in a multi-layer system of dipoles. (a) Layers are prepared in a non-coplanar spiral state with the collective layer spin pointing along $\hat{x},\hat{y},\hat{z}$ repeating periodically (gray arrows). Central layer also shows the local $x$ (red, into plane) and $y$ (blue) direction  (b) Largest singular value $\lambda_{max}$ of the spin Green's function $G^{\alpha \beta}_{i_c,j}(t)=i\langle \left[S_{i_c}^{\alpha}(t=0), S_j^{\beta}(t) \right]\rangle$ at fixed times as indicated in the legend versus distance j. Colorbar shows the (left) eigenvector structure $|v^x|^2 -  |v^y|^2$ indicating that correlations that propagate to the right originate from $S_{i_c}^x(t=0)$, while correlations that propagate to the left originate from $S_{i_c}^y(t=0)$. System with $L=10$ ($N=100$ molecules per layer) with 12 total layers.\label{fig:chirality}}
\end{figure} 

We next extend our discussion to multiple layers, and exploit the capability to prepare more complex initial states.  To this end we consider a non-coplanar spiral state where we take the layer spin directions to be $\vec{n}_{i} = \hat{x}$, $\hat{y}$ and $\hat{z}$ in order and repeating periodically along the layers (Fig.~\ref{fig:chirality}(a)). 

We then rotate the local spin basis in each layer to be aligned with the initially prepared spin direction via $S_j = U^{\dagger}_{j} \tilde{S}_j U_j$ with $U_j = e^{ i j 2 \pi/3 \left(\hat{S}_i^x + \hat{S}_i^y +\hat{S}_i^z\right)/\sqrt{3}}$.  In this rotated frame, the XXX Hamiltonian can be projected onto the fully symmetric state in each layer. If we also
restrict the interactions to nearest-layer interactions, it can be mapped to

\begin{align}
\hat{H} &= V \sum_{i} (U^{\dagger}_{i} \tilde{S}_i U_i) \cdot (U^{\dagger}_{j} \tilde{S}_j U_j)  \\
&= V \sum_i \left( \tilde{S}_i^x \tilde{S}_{i+1}^z + \tilde{S}_i^y \tilde{S}_{i+1}^x + \tilde{S}_{i}^z \tilde{S}_{i+1}^y \right)
\end{align}

We then perform a unitary transformation to remove a global rotation of all spins via $U(t) = e^{- i t V/\hbar \sum_i \left(\hat{S}_i^x + \hat{S}_i^y +\hat{S}_i^z\right)}$, and finally use a Holstein-Primakoff transformation  (\cite{supplemental}) to obtain up to  quadratic order
\begin{align}
    \hat{H} \approx S V\sum_j \left( i  \hat{a}_j \hat{a}^{\dagger}_{j+1}  - i \hat{a}^{\dagger}_j \hat{a}^{\dagger}_{j+1}  + \mathrm{h.c.} \right)
\end{align}
where $\hat{a}_j$ is a bosonic  creation operator acting on the  time and layer-dependent vacuum state. This Hamiltonian is  the bosonic version of the famous Kitaev model first introduced in Ref. \cite{Zhou_2011}.
It can be expressed in terms of hermitian quadrature operators \cite{McDonald_PhysicalReviewX_8_2018}, $ \hat{x}_j = (\hat{a}_j + \hat{a}^\dagger_j)/\sqrt{2}$, $\hat{p}_j^{\dagger} = (\hat{a}_j - \hat{a}^\dagger_j)/(i\sqrt{2}) $
as $ \hat{H}= -2 S V\sum_{j} \hat{p}_{j} \hat{x}_{j+1}$, whose  equations of motion  are fully decoupled $\dot{\hat{x}}_i =- 2 S V/\hbar \, \hat{x}_{i+1}$, and $\dot{\hat{p}}_i = 2 S V/\hbar \, \hat{p}_{i-1}$, and therefore show perfect chiral transport: the  $\hat{x}$ quadratures are only coupled to $\hat{x}$-quadratures to the right and  the  $\hat{p}$ quadratures are only coupled to $\hat{p}$-quadratures to the left.

The chiral nature of this propagation directly manifests in the corresponding Green's functions which take the following form (\cite{supplemental}) 
\begin{align}
   \left[  \hat{x}_j, \hat{p}_{j+r}(t) \right]  &=  i {(-2 S V t/\hbar)}^r/r! \, \theta(r)\\
     \left[ \hat{p}_j, \hat{x}_{j-r}(t) \right] &= -i {( 2 S V t/\hbar)}^r/r! \, \theta(r) 
\end{align}
where the step function $\theta(r)=1$ if $r\ge 0$ ensures the chiral propagation, and all other commutators vanish.  

This chiral transport can also be directly observed in the spin dynamics via the Green's function $G_{ij}^{\alpha\beta}(t) =i \langle \left[S_i^{\alpha}(t=0), S_j^{\beta}(t) \right]\rangle$. This intuitively measures the propagation of correlations to layer $j$ along the  spin component $\beta$ due to an initial infinitesimal rotation in layer $i$ around spin direction $\alpha$. Using  a short-time expansion (\cite{supplemental}), we analytically recover the chiral transport predicted by the bosonic model, i.e  for a site $i$ with the spin initially pointing along $z$ we obtain $\langle \left[\hat{S}_{i}^{x} , \hat{S}_{i+1}^z (t)\right] \rangle \approx \frac{-i V t}{\hbar} S^2$ and $\langle \left[\hat{S}_{i}^{y} , \hat{S}_{i-1}^z (t)\right] \rangle \approx \frac{-i V t}{\hbar} S^2$ indicating that $x$ perturbations propagate to the right, and $y$ propagate to the left.

We confirm this phenomenology of correlation propagation using numerical simulations of the full spin dynamics in Fig.~\ref{fig:chirality}b) which shows the largest singular value of this Green's function as a function of distance from the central layer $i_c$ chosen in such a way  that  its  collective spin initially points along $z$. When we look at the structure of the left eigenvector, which is associated to the initial rotation by $S_{i_c}^{\alpha}(t=0)$, we observe that to the right of the initial perturbation ($j>i_c$) it is (almost) purely along $x$, while to the left ($j<i_c$) it is (almost) purely along $y$. This implies that that correlations that propagate to the right originate from $S_{i_c}^x(t=0)$, while correlations that propagate to the left originate from $S_{i_c}^y(t=0)$,
demonstrating genuine  chiral behavior. We additionally compare the full dipolar spin model  to the  analytical solution of the bosonic model and observe good quantitative agreement between both solutions   up to times where boundary effects become relevant \cite{supplemental}.


{\it Outlook.---} In summary,  our work  demonstrates  the  large space of opportunities,  uniquely enabled by the capability to spatially select,  prepare and measure quantum states, to study novel non-equilibrium phenomena and to control the growth and propagation of quantum correlations  with applications in quantum sensing and simulation. While here we focused on the limit of Heisenberg  in-plane interactions which favor spin alignment  within layers, and thus homogeneous  excitations predominantly in the  fully symmetric manifold  within the layers, by  using a more general type of intralayer spin Hamiltonians  enriched by the anisotropic nature of the dipolar interactions, one  should be able to generate  spatially dependent and anisotropic excitations featuring  rich non-equilibrium universal behaviors \cite{RodriguezNieva2022}. Furthermore, the use of time-reversal protocols should enable  measurements of  out-of-time-order correlations \cite{Grttner_NatPhys_2017} to better quantify correlation growth, or to realise $SU(1,1)$ interferometry \cite{Linneman_PRL_2016}. The phenomenology discussed here might be even further enriched utilizing dipoles with a larger internal state space, e.g. multiple rotational states of molecules or larger spin magnetic atoms.

 \begin{acknowledgments}
\noindent{\textit{Acknowledgements:}
We  acknowledge careful review of this manuscript and  useful  comments  from Calder Miller, Lindsay Sonderhouse and Luis Santos.
The work  is supported by the AFOSR MURI, by  the DARPA DRINQs grant, the ARO single investigator award W911NF-19-1-0210,   the  NSF JILA-PFC PHY-1734006 grants, NSF QLCI-2016244 grants, by the DOE Quantum Systems Accelerator (QSA) grant and by NIST.

} 
.
\end{acknowledgments}


\nocite{Santos2018}

%


\begin{thebibliography}{74}%
\makeatletter
\providecommand \@ifxundefined [1]{%
 \@ifx{#1\undefined}
}%
\providecommand \@ifnum [1]{%
 \ifnum #1\expandafter \@firstoftwo
 \else \expandafter \@secondoftwo
 \fi
}%
\providecommand \@ifx [1]{%
 \ifx #1\expandafter \@firstoftwo
 \else \expandafter \@secondoftwo
 \fi
}%
\providecommand \natexlab [1]{#1}%
\providecommand \enquote  [1]{``#1''}%
\providecommand \bibnamefont  [1]{#1}%
\providecommand \bibfnamefont [1]{#1}%
\providecommand \citenamefont [1]{#1}%
\providecommand \href@noop [0]{\@secondoftwo}%
\providecommand \href [0]{\begingroup \@sanitize@url \@href}%
\providecommand \@href[1]{\@@startlink{#1}\@@href}%
\providecommand \@@href[1]{\endgroup#1\@@endlink}%
\providecommand \@sanitize@url [0]{\catcode `\\12\catcode `\$12\catcode
  `\&12\catcode `\#12\catcode `\^12\catcode `\_12\catcode `\%12\relax}%
\providecommand \@@startlink[1]{}%
\providecommand \@@endlink[0]{}%
\providecommand \url  [0]{\begingroup\@sanitize@url \@url }%
\providecommand \@url [1]{\endgroup\@href {#1}{\urlprefix }}%
\providecommand \urlprefix  [0]{URL }%
\providecommand \Eprint [0]{\href }%
\providecommand \doibase [0]{https://doi.org/}%
\providecommand \selectlanguage [0]{\@gobble}%
\providecommand \bibinfo  [0]{\@secondoftwo}%
\providecommand \bibfield  [0]{\@secondoftwo}%
\providecommand \translation [1]{[#1]}%
\providecommand \BibitemOpen [0]{}%
\providecommand \bibitemStop [0]{}%
\providecommand \bibitemNoStop [0]{.\EOS\space}%
\providecommand \EOS [0]{\spacefactor3000\relax}%
\providecommand \BibitemShut  [1]{\csname bibitem#1\endcsname}%
\let\auto@bib@innerbib\@empty
\bibitem [{\citenamefont {Gross}\ and\ \citenamefont
  {Bakr}(2021)}]{NatPhysGross2021}%
  \BibitemOpen
  \bibfield  {author} {\bibinfo {author} {\bibfnamefont {C.}~\bibnamefont
  {Gross}}\ and\ \bibinfo {author} {\bibfnamefont {W.~S.}\ \bibnamefont
  {Bakr}},\ }\bibfield  {title} {\bibinfo {title} {Quantum gas microscopy for
  single atom and spin detection},\ }\href
  {https://doi.org/10.1038/s41567-021-01370-5} {\bibfield  {journal} {\bibinfo
  {journal} {Nature Physics}\ }\textbf {\bibinfo {volume} {17}},\ \bibinfo
  {pages} {1316} (\bibinfo {year} {2021})}\BibitemShut {NoStop}%
\bibitem [{\citenamefont {Kaufman}\ and\ \citenamefont
  {Ni}(2021)}]{NatPhysKaufman2021}%
  \BibitemOpen
  \bibfield  {author} {\bibinfo {author} {\bibfnamefont {A.~M.}\ \bibnamefont
  {Kaufman}}\ and\ \bibinfo {author} {\bibfnamefont {K.-K.}\ \bibnamefont
  {Ni}},\ }\bibfield  {title} {\bibinfo {title} {Quantum science with optical
  tweezer arrays of ultracold atoms and molecules},\ }\href
  {https://doi.org/10.1038/s41567-021-01357-2} {\bibfield  {journal} {\bibinfo
  {journal} {Nature Physics}\ }\textbf {\bibinfo {volume} {17}},\ \bibinfo
  {pages} {1324} (\bibinfo {year} {2021})}\BibitemShut {NoStop}%
\bibitem [{\citenamefont {Vale}\ and\ \citenamefont
  {Zwierlein}(2021)}]{NatPhysVale2021}%
  \BibitemOpen
  \bibfield  {author} {\bibinfo {author} {\bibfnamefont {C.~J.}\ \bibnamefont
  {Vale}}\ and\ \bibinfo {author} {\bibfnamefont {M.}~\bibnamefont
  {Zwierlein}},\ }\bibfield  {title} {\bibinfo {title} {Spectroscopic probes of
  quantum gases},\ }\href {https://doi.org/10.1038/s41567-021-01434-6}
  {\bibfield  {journal} {\bibinfo  {journal} {Nature Physics}\ }\textbf
  {\bibinfo {volume} {17}},\ \bibinfo {pages} {1305} (\bibinfo {year}
  {2021})}\BibitemShut {NoStop}%
\bibitem [{\citenamefont {Bloch}(2005)}]{NPhysBloch2005}%
  \BibitemOpen
  \bibfield  {author} {\bibinfo {author} {\bibfnamefont {I.}~\bibnamefont
  {Bloch}},\ }\bibfield  {title} {\bibinfo {title} {Ultracold quantum gases in
  optical lattices},\ }\href {https://doi.org/10.1038/nphys138} {\bibfield
  {journal} {\bibinfo  {journal} {Nature Physics}\ }\textbf {\bibinfo {volume}
  {1}},\ \bibinfo {pages} {23} (\bibinfo {year} {2005})}\BibitemShut {NoStop}%
\bibitem [{\citenamefont {Bloch}\ \emph {et~al.}(2008)\citenamefont {Bloch},
  \citenamefont {Dalibard},\ and\ \citenamefont {Zwerger}}]{RevModPhys.80.885}%
  \BibitemOpen
  \bibfield  {author} {\bibinfo {author} {\bibfnamefont {I.}~\bibnamefont
  {Bloch}}, \bibinfo {author} {\bibfnamefont {J.}~\bibnamefont {Dalibard}},\
  and\ \bibinfo {author} {\bibfnamefont {W.}~\bibnamefont {Zwerger}},\
  }\bibfield  {title} {\bibinfo {title} {Many-body physics with ultracold
  gases},\ }\href {https://doi.org/10.1103/RevModPhys.80.885} {\bibfield
  {journal} {\bibinfo  {journal} {Rev. Mod. Phys.}\ }\textbf {\bibinfo {volume}
  {80}},\ \bibinfo {pages} {885} (\bibinfo {year} {2008})}\BibitemShut
  {NoStop}%
\bibitem [{\citenamefont {Bohn}\ \emph {et~al.}(2017)\citenamefont {Bohn},
  \citenamefont {Rey},\ and\ \citenamefont {Ye}}]{Bohn_Science_357_2017}%
  \BibitemOpen
  \bibfield  {author} {\bibinfo {author} {\bibfnamefont {J.~L.}\ \bibnamefont
  {Bohn}}, \bibinfo {author} {\bibfnamefont {A.~M.}\ \bibnamefont {Rey}},\ and\
  \bibinfo {author} {\bibfnamefont {J.}~\bibnamefont {Ye}},\ }\bibfield
  {title} {\bibinfo {title} {Cold molecules: Progress in quantum engineering of
  chemistry and quantum matter},\ }\href
  {https://doi.org/10.1126/science.aam6299} {\bibfield  {journal} {\bibinfo
  {journal} {Science}\ }\textbf {\bibinfo {volume} {357}},\ \bibinfo {pages}
  {1002} (\bibinfo {year} {2017})}\BibitemShut {NoStop}%
\bibitem [{\citenamefont {Moses}\ \emph {et~al.}(2016)\citenamefont {Moses},
  \citenamefont {Covey}, \citenamefont {Miecnikowski}, \citenamefont {Jin},\
  and\ \citenamefont {Ye}}]{Moses2016}%
  \BibitemOpen
  \bibfield  {author} {\bibinfo {author} {\bibfnamefont {S.~A.}\ \bibnamefont
  {Moses}}, \bibinfo {author} {\bibfnamefont {J.~P.}\ \bibnamefont {Covey}},
  \bibinfo {author} {\bibfnamefont {M.~T.}\ \bibnamefont {Miecnikowski}},
  \bibinfo {author} {\bibfnamefont {D.~S.}\ \bibnamefont {Jin}},\ and\ \bibinfo
  {author} {\bibfnamefont {J.}~\bibnamefont {Ye}},\ }\bibfield  {title}
  {\bibinfo {title} {New frontiers for quantum gases of polar molecules},\
  }\href {https://doi.org/10.1038/nphys3985} {\bibfield  {journal} {\bibinfo
  {journal} {Nature Physics}\ }\textbf {\bibinfo {volume} {13}},\ \bibinfo
  {pages} {13} (\bibinfo {year} {2016})}\BibitemShut {NoStop}%
\bibitem [{\citenamefont {Baranov}\ \emph {et~al.}(2012)\citenamefont
  {Baranov}, \citenamefont {Dalmonte}, \citenamefont {Pupillo},\ and\
  \citenamefont {Zoller}}]{Baranov_ChemicalReviews_112_2012}%
  \BibitemOpen
  \bibfield  {author} {\bibinfo {author} {\bibfnamefont {M.~A.}\ \bibnamefont
  {Baranov}}, \bibinfo {author} {\bibfnamefont {M.}~\bibnamefont {Dalmonte}},
  \bibinfo {author} {\bibfnamefont {G.}~\bibnamefont {Pupillo}},\ and\ \bibinfo
  {author} {\bibfnamefont {P.}~\bibnamefont {Zoller}},\ }\bibfield  {title}
  {\bibinfo {title} {Condensed matter theory of dipolar quantum gases},\ }\href
  {https://doi.org/10.1021/cr2003568} {\bibfield  {journal} {\bibinfo
  {journal} {Chemical Reviews}\ }\textbf {\bibinfo {volume} {112}},\ \bibinfo
  {pages} {5012} (\bibinfo {year} {2012})}\BibitemShut {NoStop}%
\bibitem [{\citenamefont {Browaeys}\ and\ \citenamefont
  {Lahaye}(2020)}]{NatPhysBrowaeys2020}%
  \BibitemOpen
  \bibfield  {author} {\bibinfo {author} {\bibfnamefont {A.}~\bibnamefont
  {Browaeys}}\ and\ \bibinfo {author} {\bibfnamefont {T.}~\bibnamefont
  {Lahaye}},\ }\bibfield  {title} {\bibinfo {title} {Many-body physics with
  individually controlled rydberg atoms},\ }\href
  {https://doi.org/10.1038/s41567-019-0733-z} {\bibfield  {journal} {\bibinfo
  {journal} {Nature Physics}\ }\textbf {\bibinfo {volume} {16}},\ \bibinfo
  {pages} {132} (\bibinfo {year} {2020})}\BibitemShut {NoStop}%
\bibitem [{\citenamefont {Morgado}\ and\ \citenamefont
  {Whitlock}(2021)}]{QScMorgado2021}%
  \BibitemOpen
  \bibfield  {author} {\bibinfo {author} {\bibfnamefont {M.}~\bibnamefont
  {Morgado}}\ and\ \bibinfo {author} {\bibfnamefont {S.}~\bibnamefont
  {Whitlock}},\ }\bibfield  {title} {\bibinfo {title} {Quantum simulation and
  computing with rydberg-interacting qubits},\ }\href
  {https://doi.org/10.1116/5.0036562} {\bibfield  {journal} {\bibinfo
  {journal} {{AVS} Quantum Science}\ }\textbf {\bibinfo {volume} {3}},\
  \bibinfo {pages} {023501} (\bibinfo {year} {2021})}\BibitemShut {NoStop}%
\bibitem [{\citenamefont {Lahaye}\ \emph {et~al.}(2009)\citenamefont {Lahaye},
  \citenamefont {Menotti}, \citenamefont {Santos}, \citenamefont {Lewenstein},\
  and\ \citenamefont {Pfau}}]{RepProgPhysLahaye2009}%
  \BibitemOpen
  \bibfield  {author} {\bibinfo {author} {\bibfnamefont {T.}~\bibnamefont
  {Lahaye}}, \bibinfo {author} {\bibfnamefont {C.}~\bibnamefont {Menotti}},
  \bibinfo {author} {\bibfnamefont {L.}~\bibnamefont {Santos}}, \bibinfo
  {author} {\bibfnamefont {M.}~\bibnamefont {Lewenstein}},\ and\ \bibinfo
  {author} {\bibfnamefont {T.}~\bibnamefont {Pfau}},\ }\bibfield  {title}
  {\bibinfo {title} {The physics of dipolar bosonic quantum gases},\ }\href
  {https://doi.org/10.1088/0034-4885/72/12/126401} {\bibfield  {journal}
  {\bibinfo  {journal} {Reports on Progress in Physics}\ }\textbf {\bibinfo
  {volume} {72}},\ \bibinfo {pages} {126401} (\bibinfo {year}
  {2009})}\BibitemShut {NoStop}%
\bibitem [{\citenamefont {{Chomaz}}\ \emph {et~al.}(2022)\citenamefont
  {{Chomaz}}, \citenamefont {{Ferrier-Barbut}}, \citenamefont {{Ferlaino}},
  \citenamefont {{Laburthe-Tolra}}, \citenamefont {{Lev}},\ and\ \citenamefont
  {{Pfau}}}]{Chomaz2022}%
  \BibitemOpen
  \bibfield  {author} {\bibinfo {author} {\bibfnamefont {L.}~\bibnamefont
  {{Chomaz}}}, \bibinfo {author} {\bibfnamefont {I.}~\bibnamefont
  {{Ferrier-Barbut}}}, \bibinfo {author} {\bibfnamefont {F.}~\bibnamefont
  {{Ferlaino}}}, \bibinfo {author} {\bibfnamefont {B.}~\bibnamefont
  {{Laburthe-Tolra}}}, \bibinfo {author} {\bibfnamefont {B.~L.}\ \bibnamefont
  {{Lev}}},\ and\ \bibinfo {author} {\bibfnamefont {T.}~\bibnamefont
  {{Pfau}}},\ }\bibfield  {title} {\bibinfo {title} {{Dipolar physics: A review
  of experiments with magnetic quantum gases}},\ }\href@noop {} {\bibfield
  {journal} {\bibinfo  {journal} {arXiv e-prints}\ ,\ \bibinfo {eid}
  {arXiv:2201.02672}} (\bibinfo {year} {2022})},\ \Eprint
  {https://arxiv.org/abs/2201.02672} {arXiv:2201.02672 [cond-mat.quant-gas]}
  \BibitemShut {NoStop}%
\bibitem [{\citenamefont {{Defenu}}\ \emph {et~al.}(2021)\citenamefont
  {{Defenu}}, \citenamefont {{Donner}}, \citenamefont {{Macr{\`\i}}},
  \citenamefont {{Pagano}}, \citenamefont {{Ruffo}},\ and\ \citenamefont
  {{Trombettoni}}}]{ArxivDefenu2021}%
  \BibitemOpen
  \bibfield  {author} {\bibinfo {author} {\bibfnamefont {N.}~\bibnamefont
  {{Defenu}}}, \bibinfo {author} {\bibfnamefont {T.}~\bibnamefont {{Donner}}},
  \bibinfo {author} {\bibfnamefont {T.}~\bibnamefont {{Macr{\`\i}}}}, \bibinfo
  {author} {\bibfnamefont {G.}~\bibnamefont {{Pagano}}}, \bibinfo {author}
  {\bibfnamefont {S.}~\bibnamefont {{Ruffo}}},\ and\ \bibinfo {author}
  {\bibfnamefont {A.}~\bibnamefont {{Trombettoni}}},\ }\bibfield  {title}
  {\bibinfo {title} {{Long-range interacting quantum systems}},\ }\href@noop {}
  {\bibfield  {journal} {\bibinfo  {journal} {arXiv e-prints}\ ,\ \bibinfo
  {eid} {arXiv:2109.01063}} (\bibinfo {year} {2021})},\ \Eprint
  {https://arxiv.org/abs/2109.01063} {arXiv:2109.01063 [cond-mat.quant-gas]}
  \BibitemShut {NoStop}%
\bibitem [{\citenamefont {Mivehvar}\ \emph {et~al.}(2021)\citenamefont
  {Mivehvar}, \citenamefont {Piazza}, \citenamefont {Donner},\ and\
  \citenamefont {Ritsch}}]{AiPMivehvar2021}%
  \BibitemOpen
  \bibfield  {author} {\bibinfo {author} {\bibfnamefont {F.}~\bibnamefont
  {Mivehvar}}, \bibinfo {author} {\bibfnamefont {F.}~\bibnamefont {Piazza}},
  \bibinfo {author} {\bibfnamefont {T.}~\bibnamefont {Donner}},\ and\ \bibinfo
  {author} {\bibfnamefont {H.}~\bibnamefont {Ritsch}},\ }\bibfield  {title}
  {\bibinfo {title} {Cavity {QED} with quantum gases: new paradigms in
  many-body physics},\ }\href {https://doi.org/10.1080/00018732.2021.1969727}
  {\bibfield  {journal} {\bibinfo  {journal} {Advances in Physics}\ }\textbf
  {\bibinfo {volume} {70}},\ \bibinfo {pages} {1} (\bibinfo {year}
  {2021})}\BibitemShut {NoStop}%
\bibitem [{\citenamefont {Norcia}\ \emph {et~al.}(2018)\citenamefont {Norcia},
  \citenamefont {Lewis-Swan}, \citenamefont {Cline}, \citenamefont {Zhu},
  \citenamefont {Rey},\ and\ \citenamefont {Thompson}}]{ScienceNorcia2018}%
  \BibitemOpen
  \bibfield  {author} {\bibinfo {author} {\bibfnamefont {M.~A.}\ \bibnamefont
  {Norcia}}, \bibinfo {author} {\bibfnamefont {R.~J.}\ \bibnamefont
  {Lewis-Swan}}, \bibinfo {author} {\bibfnamefont {J.~R.~K.}\ \bibnamefont
  {Cline}}, \bibinfo {author} {\bibfnamefont {B.}~\bibnamefont {Zhu}}, \bibinfo
  {author} {\bibfnamefont {A.~M.}\ \bibnamefont {Rey}},\ and\ \bibinfo {author}
  {\bibfnamefont {J.~K.}\ \bibnamefont {Thompson}},\ }\bibfield  {title}
  {\bibinfo {title} {Cavity-mediated collective spin-exchange interactions in a
  strontium superradiant laser},\ }\href
  {https://doi.org/10.1126/science.aar3102} {\bibfield  {journal} {\bibinfo
  {journal} {Science}\ }\textbf {\bibinfo {volume} {361}},\ \bibinfo {pages}
  {259} (\bibinfo {year} {2018})}\BibitemShut {NoStop}%
\bibitem [{\citenamefont {Masson}\ \emph {et~al.}(2017)\citenamefont {Masson},
  \citenamefont {Barrett},\ and\ \citenamefont {Parkins}}]{PRLParkings2017}%
  \BibitemOpen
  \bibfield  {author} {\bibinfo {author} {\bibfnamefont {S.~J.}\ \bibnamefont
  {Masson}}, \bibinfo {author} {\bibfnamefont {M.~D.}\ \bibnamefont
  {Barrett}},\ and\ \bibinfo {author} {\bibfnamefont {S.}~\bibnamefont
  {Parkins}},\ }\bibfield  {title} {\bibinfo {title} {Cavity qed engineering of
  spin dynamics and squeezing in a spinor gas},\ }\href
  {https://doi.org/10.1103/PhysRevLett.119.213601} {\bibfield  {journal}
  {\bibinfo  {journal} {Phys. Rev. Lett.}\ }\textbf {\bibinfo {volume} {119}},\
  \bibinfo {pages} {213601} (\bibinfo {year} {2017})}\BibitemShut {NoStop}%
\bibitem [{\citenamefont {Davis}\ \emph {et~al.}(2019)\citenamefont {Davis},
  \citenamefont {Bentsen}, \citenamefont {Homeier}, \citenamefont {Li},\ and\
  \citenamefont {Schleier-Smith}}]{PRLSchleierSmith2019}%
  \BibitemOpen
  \bibfield  {author} {\bibinfo {author} {\bibfnamefont {E.~J.}\ \bibnamefont
  {Davis}}, \bibinfo {author} {\bibfnamefont {G.}~\bibnamefont {Bentsen}},
  \bibinfo {author} {\bibfnamefont {L.}~\bibnamefont {Homeier}}, \bibinfo
  {author} {\bibfnamefont {T.}~\bibnamefont {Li}},\ and\ \bibinfo {author}
  {\bibfnamefont {M.~H.}\ \bibnamefont {Schleier-Smith}},\ }\bibfield  {title}
  {\bibinfo {title} {Photon-mediated spin-exchange dynamics of spin-1 atoms},\
  }\href {https://doi.org/10.1103/PhysRevLett.122.010405} {\bibfield  {journal}
  {\bibinfo  {journal} {Phys. Rev. Lett.}\ }\textbf {\bibinfo {volume} {122}},\
  \bibinfo {pages} {010405} (\bibinfo {year} {2019})}\BibitemShut {NoStop}%
\bibitem [{\citenamefont {Blatt}\ and\ \citenamefont
  {Roos}(2012)}]{NatPhysBlatt2012}%
  \BibitemOpen
  \bibfield  {author} {\bibinfo {author} {\bibfnamefont {R.}~\bibnamefont
  {Blatt}}\ and\ \bibinfo {author} {\bibfnamefont {C.~F.}\ \bibnamefont
  {Roos}},\ }\bibfield  {title} {\bibinfo {title} {Quantum simulations with
  trapped ions},\ }\href {https://doi.org/10.1038/nphys2252} {\bibfield
  {journal} {\bibinfo  {journal} {Nature Physics}\ }\textbf {\bibinfo {volume}
  {8}},\ \bibinfo {pages} {277} (\bibinfo {year} {2012})}\BibitemShut {NoStop}%
\bibitem [{\citenamefont {Monroe}\ \emph {et~al.}(2021)\citenamefont {Monroe},
  \citenamefont {Campbell}, \citenamefont {Duan}, \citenamefont {Gong},
  \citenamefont {Gorshkov}, \citenamefont {Hess}, \citenamefont {Islam},
  \citenamefont {Kim}, \citenamefont {Linke}, \citenamefont {Pagano},
  \citenamefont {Richerme}, \citenamefont {Senko},\ and\ \citenamefont
  {Yao}}]{RevModMonroe2021}%
  \BibitemOpen
  \bibfield  {author} {\bibinfo {author} {\bibfnamefont {C.}~\bibnamefont
  {Monroe}}, \bibinfo {author} {\bibfnamefont {W.~C.}\ \bibnamefont
  {Campbell}}, \bibinfo {author} {\bibfnamefont {L.-M.}\ \bibnamefont {Duan}},
  \bibinfo {author} {\bibfnamefont {Z.-X.}\ \bibnamefont {Gong}}, \bibinfo
  {author} {\bibfnamefont {A.~V.}\ \bibnamefont {Gorshkov}}, \bibinfo {author}
  {\bibfnamefont {P.~W.}\ \bibnamefont {Hess}}, \bibinfo {author}
  {\bibfnamefont {R.}~\bibnamefont {Islam}}, \bibinfo {author} {\bibfnamefont
  {K.}~\bibnamefont {Kim}}, \bibinfo {author} {\bibfnamefont {N.~M.}\
  \bibnamefont {Linke}}, \bibinfo {author} {\bibfnamefont {G.}~\bibnamefont
  {Pagano}}, \bibinfo {author} {\bibfnamefont {P.}~\bibnamefont {Richerme}},
  \bibinfo {author} {\bibfnamefont {C.}~\bibnamefont {Senko}},\ and\ \bibinfo
  {author} {\bibfnamefont {N.~Y.}\ \bibnamefont {Yao}},\ }\bibfield  {title}
  {\bibinfo {title} {Programmable quantum simulations of spin systems with
  trapped ions},\ }\href {https://doi.org/10.1103/revmodphys.93.025001}
  {\bibfield  {journal} {\bibinfo  {journal} {Reviews of Modern Physics}\
  }\textbf {\bibinfo {volume} {93}},\ \bibinfo {pages} {025001} (\bibinfo
  {year} {2021})}\BibitemShut {NoStop}%
\bibitem [{\citenamefont {Bloch}\ \emph {et~al.}(2012)\citenamefont {Bloch},
  \citenamefont {Dalibard},\ and\ \citenamefont
  {Nascimb{\`{e}}ne}}]{NPhysBloch2012}%
  \BibitemOpen
  \bibfield  {author} {\bibinfo {author} {\bibfnamefont {I.}~\bibnamefont
  {Bloch}}, \bibinfo {author} {\bibfnamefont {J.}~\bibnamefont {Dalibard}},\
  and\ \bibinfo {author} {\bibfnamefont {S.}~\bibnamefont {Nascimb{\`{e}}ne}},\
  }\bibfield  {title} {\bibinfo {title} {Quantum simulations with ultracold
  quantum gases},\ }\href {https://doi.org/10.1038/nphys2259} {\bibfield
  {journal} {\bibinfo  {journal} {Nature Physics}\ }\textbf {\bibinfo {volume}
  {8}},\ \bibinfo {pages} {267} (\bibinfo {year} {2012})}\BibitemShut {NoStop}%
\bibitem [{\citenamefont {Gross}\ and\ \citenamefont
  {Bloch}(2017)}]{ScienceGross2017}%
  \BibitemOpen
  \bibfield  {author} {\bibinfo {author} {\bibfnamefont {C.}~\bibnamefont
  {Gross}}\ and\ \bibinfo {author} {\bibfnamefont {I.}~\bibnamefont {Bloch}},\
  }\bibfield  {title} {\bibinfo {title} {Quantum simulations with ultracold
  atoms in optical lattices},\ }\href {https://doi.org/10.1126/science.aal3837}
  {\bibfield  {journal} {\bibinfo  {journal} {Science}\ }\textbf {\bibinfo
  {volume} {357}},\ \bibinfo {pages} {995} (\bibinfo {year}
  {2017})}\BibitemShut {NoStop}%
\bibitem [{\citenamefont {Daley}\ \emph {et~al.}(2022)\citenamefont {Daley},
  \citenamefont {Bloch}, \citenamefont {Kokail}, \citenamefont {Flannigan},
  \citenamefont {Pearson}, \citenamefont {Troyer},\ and\ \citenamefont
  {Zoller}}]{NatureDaley2022}%
  \BibitemOpen
  \bibfield  {author} {\bibinfo {author} {\bibfnamefont {A.~J.}\ \bibnamefont
  {Daley}}, \bibinfo {author} {\bibfnamefont {I.}~\bibnamefont {Bloch}},
  \bibinfo {author} {\bibfnamefont {C.}~\bibnamefont {Kokail}}, \bibinfo
  {author} {\bibfnamefont {S.}~\bibnamefont {Flannigan}}, \bibinfo {author}
  {\bibfnamefont {N.}~\bibnamefont {Pearson}}, \bibinfo {author} {\bibfnamefont
  {M.}~\bibnamefont {Troyer}},\ and\ \bibinfo {author} {\bibfnamefont
  {P.}~\bibnamefont {Zoller}},\ }\bibfield  {title} {\bibinfo {title}
  {Practical quantum advantage in quantum simulation},\ }\href
  {https://doi.org/10.1038/s41586-022-04940-6} {\bibfield  {journal} {\bibinfo
  {journal} {Nature}\ }\textbf {\bibinfo {volume} {607}},\ \bibinfo {pages}
  {667} (\bibinfo {year} {2022})}\BibitemShut {NoStop}%
\bibitem [{\citenamefont {Pezz\`e}\ \emph {et~al.}(2018)\citenamefont
  {Pezz\`e}, \citenamefont {Smerzi}, \citenamefont {Oberthaler}, \citenamefont
  {Schmied},\ and\ \citenamefont {Treutlein}}]{RevModPhys.90.035005}%
  \BibitemOpen
  \bibfield  {author} {\bibinfo {author} {\bibfnamefont {L.}~\bibnamefont
  {Pezz\`e}}, \bibinfo {author} {\bibfnamefont {A.}~\bibnamefont {Smerzi}},
  \bibinfo {author} {\bibfnamefont {M.~K.}\ \bibnamefont {Oberthaler}},
  \bibinfo {author} {\bibfnamefont {R.}~\bibnamefont {Schmied}},\ and\ \bibinfo
  {author} {\bibfnamefont {P.}~\bibnamefont {Treutlein}},\ }\bibfield  {title}
  {\bibinfo {title} {Quantum metrology with nonclassical states of atomic
  ensembles},\ }\href {https://doi.org/10.1103/RevModPhys.90.035005} {\bibfield
   {journal} {\bibinfo  {journal} {Rev. Mod. Phys.}\ }\textbf {\bibinfo
  {volume} {90}},\ \bibinfo {pages} {035005} (\bibinfo {year}
  {2018})}\BibitemShut {NoStop}%
\bibitem [{\citenamefont {Briegel}\ \emph {et~al.}(2000)\citenamefont
  {Briegel}, \citenamefont {Calarco}, \citenamefont {Jaksch}, \citenamefont
  {Cirac},\ and\ \citenamefont {Zoller}}]{Briegel_2000}%
  \BibitemOpen
  \bibfield  {author} {\bibinfo {author} {\bibfnamefont {H.-J.}\ \bibnamefont
  {Briegel}}, \bibinfo {author} {\bibfnamefont {T.}~\bibnamefont {Calarco}},
  \bibinfo {author} {\bibfnamefont {D.}~\bibnamefont {Jaksch}}, \bibinfo
  {author} {\bibfnamefont {J.~I.}\ \bibnamefont {Cirac}},\ and\ \bibinfo
  {author} {\bibfnamefont {P.}~\bibnamefont {Zoller}},\ }\bibfield  {title}
  {\bibinfo {title} {Quantum computing with neutral atoms},\ }\href
  {https://doi.org/10.1080/09500340008244052} {\bibfield  {journal} {\bibinfo
  {journal} {Journal of Modern Optics}\ }\textbf {\bibinfo {volume} {47}},\
  \bibinfo {pages} {415} (\bibinfo {year} {2000})}\BibitemShut {NoStop}%
\bibitem [{\citenamefont {Weiss}\ and\ \citenamefont
  {Saffman}(2017)}]{Weiss2017}%
  \BibitemOpen
  \bibfield  {author} {\bibinfo {author} {\bibfnamefont {D.~S.}\ \bibnamefont
  {Weiss}}\ and\ \bibinfo {author} {\bibfnamefont {M.}~\bibnamefont
  {Saffman}},\ }\bibfield  {title} {\bibinfo {title} {Quantum computing with
  neutral atoms},\ }\href {https://doi.org/10.1063/PT.3.3626} {\bibfield
  {journal} {\bibinfo  {journal} {Physics Today}\ }\textbf {\bibinfo {volume}
  {70}},\ \bibinfo {pages} {44} (\bibinfo {year} {2017})}\BibitemShut {NoStop}%
\bibitem [{\citenamefont {Henriet}\ \emph {et~al.}(2020)\citenamefont
  {Henriet}, \citenamefont {Beguin}, \citenamefont {Signoles}, \citenamefont
  {Lahaye}, \citenamefont {Browaeys}, \citenamefont {Reymond},\ and\
  \citenamefont {Jurczak}}]{Henriet2020quantumcomputing}%
  \BibitemOpen
  \bibfield  {author} {\bibinfo {author} {\bibfnamefont {L.}~\bibnamefont
  {Henriet}}, \bibinfo {author} {\bibfnamefont {L.}~\bibnamefont {Beguin}},
  \bibinfo {author} {\bibfnamefont {A.}~\bibnamefont {Signoles}}, \bibinfo
  {author} {\bibfnamefont {T.}~\bibnamefont {Lahaye}}, \bibinfo {author}
  {\bibfnamefont {A.}~\bibnamefont {Browaeys}}, \bibinfo {author}
  {\bibfnamefont {G.-O.}\ \bibnamefont {Reymond}},\ and\ \bibinfo {author}
  {\bibfnamefont {C.}~\bibnamefont {Jurczak}},\ }\bibfield  {title} {\bibinfo
  {title} {Quantum computing with neutral atoms},\ }\href
  {https://doi.org/10.22331/q-2020-09-21-327} {\bibfield  {journal} {\bibinfo
  {journal} {{Quantum}}\ }\textbf {\bibinfo {volume} {4}},\ \bibinfo {pages}
  {327} (\bibinfo {year} {2020})}\BibitemShut {NoStop}%
\bibitem [{\citenamefont {Lewis-Swan}\ \emph {et~al.}(2019)\citenamefont
  {Lewis-Swan}, \citenamefont {Safavi-Naini}, \citenamefont {Kaufman},\ and\
  \citenamefont {Rey}}]{NatRevPhys_Lewis_Swan_2019}%
  \BibitemOpen
  \bibfield  {author} {\bibinfo {author} {\bibfnamefont {R.~J.}\ \bibnamefont
  {Lewis-Swan}}, \bibinfo {author} {\bibfnamefont {A.}~\bibnamefont
  {Safavi-Naini}}, \bibinfo {author} {\bibfnamefont {A.~M.}\ \bibnamefont
  {Kaufman}},\ and\ \bibinfo {author} {\bibfnamefont {A.~M.}\ \bibnamefont
  {Rey}},\ }\bibfield  {title} {\bibinfo {title} {Dynamics of quantum
  information},\ }\href {https://doi.org/10.1038/s42254-019-0090-y} {\bibfield
  {journal} {\bibinfo  {journal} {Nature Reviews Physics}\ }\textbf {\bibinfo
  {volume} {1}},\ \bibinfo {pages} {627} (\bibinfo {year} {2019})}\BibitemShut
  {NoStop}%
\bibitem [{\citenamefont {Einstein}\ \emph {et~al.}(1935)\citenamefont
  {Einstein}, \citenamefont {Podolsky},\ and\ \citenamefont
  {Rosen}}]{EPR_1935}%
  \BibitemOpen
  \bibfield  {author} {\bibinfo {author} {\bibfnamefont {A.}~\bibnamefont
  {Einstein}}, \bibinfo {author} {\bibfnamefont {B.}~\bibnamefont {Podolsky}},\
  and\ \bibinfo {author} {\bibfnamefont {N.}~\bibnamefont {Rosen}},\ }\bibfield
   {title} {\bibinfo {title} {Can quantum-mechanical description of physical
  reality be considered complete?},\ }\href
  {https://doi.org/10.1103/PhysRev.47.777} {\bibfield  {journal} {\bibinfo
  {journal} {Phys. Rev.}\ }\textbf {\bibinfo {volume} {47}},\ \bibinfo {pages}
  {777} (\bibinfo {year} {1935})}\BibitemShut {NoStop}%
\bibitem [{\citenamefont {Reid}\ \emph {et~al.}(2009)\citenamefont {Reid},
  \citenamefont {Drummond}, \citenamefont {Bowen}, \citenamefont {Cavalcanti},
  \citenamefont {Lam}, \citenamefont {Bachor}, \citenamefont {Andersen},\ and\
  \citenamefont {Leuchs}}]{RevModPhys.81.1727}%
  \BibitemOpen
  \bibfield  {author} {\bibinfo {author} {\bibfnamefont {M.~D.}\ \bibnamefont
  {Reid}}, \bibinfo {author} {\bibfnamefont {P.~D.}\ \bibnamefont {Drummond}},
  \bibinfo {author} {\bibfnamefont {W.~P.}\ \bibnamefont {Bowen}}, \bibinfo
  {author} {\bibfnamefont {E.~G.}\ \bibnamefont {Cavalcanti}}, \bibinfo
  {author} {\bibfnamefont {P.~K.}\ \bibnamefont {Lam}}, \bibinfo {author}
  {\bibfnamefont {H.~A.}\ \bibnamefont {Bachor}}, \bibinfo {author}
  {\bibfnamefont {U.~L.}\ \bibnamefont {Andersen}},\ and\ \bibinfo {author}
  {\bibfnamefont {G.}~\bibnamefont {Leuchs}},\ }\bibfield  {title} {\bibinfo
  {title} {Colloquium: The einstein-podolsky-rosen paradox: From concepts to
  applications},\ }\href {https://doi.org/10.1103/RevModPhys.81.1727}
  {\bibfield  {journal} {\bibinfo  {journal} {Rev. Mod. Phys.}\ }\textbf
  {\bibinfo {volume} {81}},\ \bibinfo {pages} {1727} (\bibinfo {year}
  {2009})}\BibitemShut {NoStop}%
\bibitem [{\citenamefont {Agarwal}(2013)}]{agarwal2013quantum}%
  \BibitemOpen
  \bibfield  {author} {\bibinfo {author} {\bibfnamefont {G.}~\bibnamefont
  {Agarwal}},\ }\href {https://books.google.com/books?id=7KKw\_XIYaioC} {\emph
  {\bibinfo {title} {Quantum Optics}}},\ Quantum Optics\ (\bibinfo  {publisher}
  {Cambridge University Press},\ \bibinfo {year} {2013})\BibitemShut {NoStop}%
\bibitem [{\citenamefont {Caves}\ and\ \citenamefont
  {Schumaker}(1985)}]{Caves1985}%
  \BibitemOpen
  \bibfield  {author} {\bibinfo {author} {\bibfnamefont {C.~M.}\ \bibnamefont
  {Caves}}\ and\ \bibinfo {author} {\bibfnamefont {B.~L.}\ \bibnamefont
  {Schumaker}},\ }\bibfield  {title} {\bibinfo {title} {New formalism for
  two-photon quantum optics. {I}. quadrature phases and squeezed states},\
  }\href {https://doi.org/10.1103/physreva.31.3068} {\bibfield  {journal}
  {\bibinfo  {journal} {Physical Review A}\ }\textbf {\bibinfo {volume} {31}},\
  \bibinfo {pages} {3068} (\bibinfo {year} {1985})}\BibitemShut {NoStop}%
\bibitem [{\citenamefont {Schumaker}\ and\ \citenamefont
  {Caves}(1985)}]{Schumaker1985}%
  \BibitemOpen
  \bibfield  {author} {\bibinfo {author} {\bibfnamefont {B.~L.}\ \bibnamefont
  {Schumaker}}\ and\ \bibinfo {author} {\bibfnamefont {C.~M.}\ \bibnamefont
  {Caves}},\ }\bibfield  {title} {\bibinfo {title} {New formalism for
  two-photon quantum optics. {II}. mathematical foundation and compact
  notation},\ }\href {https://doi.org/10.1103/physreva.31.3093} {\bibfield
  {journal} {\bibinfo  {journal} {Physical Review A}\ }\textbf {\bibinfo
  {volume} {31}},\ \bibinfo {pages} {3093} (\bibinfo {year}
  {1985})}\BibitemShut {NoStop}%
\bibitem [{\citenamefont {Hauke}\ \emph {et~al.}(2013)\citenamefont {Hauke},
  \citenamefont {Marcos}, \citenamefont {Dalmonte},\ and\ \citenamefont
  {Zoller}}]{Hauke_PhysRevX_2013}%
  \BibitemOpen
  \bibfield  {author} {\bibinfo {author} {\bibfnamefont {P.}~\bibnamefont
  {Hauke}}, \bibinfo {author} {\bibfnamefont {D.}~\bibnamefont {Marcos}},
  \bibinfo {author} {\bibfnamefont {M.}~\bibnamefont {Dalmonte}},\ and\
  \bibinfo {author} {\bibfnamefont {P.}~\bibnamefont {Zoller}},\ }\bibfield
  {title} {\bibinfo {title} {Quantum simulation of a lattice schwinger model in
  a chain of trapped ions},\ }\href {https://doi.org/10.1103/PhysRevX.3.041018}
  {\bibfield  {journal} {\bibinfo  {journal} {Phys. Rev. X}\ }\textbf {\bibinfo
  {volume} {3}},\ \bibinfo {pages} {041018} (\bibinfo {year}
  {2013})}\BibitemShut {NoStop}%
\bibitem [{\citenamefont {Kasper}\ \emph {et~al.}(2016)\citenamefont {Kasper},
  \citenamefont {Hebenstreit}, \citenamefont {Oberthaler},\ and\ \citenamefont
  {Berges}}]{Kasper_PhysLettB_2016}%
  \BibitemOpen
  \bibfield  {author} {\bibinfo {author} {\bibfnamefont {V.}~\bibnamefont
  {Kasper}}, \bibinfo {author} {\bibfnamefont {F.}~\bibnamefont {Hebenstreit}},
  \bibinfo {author} {\bibfnamefont {M.}~\bibnamefont {Oberthaler}},\ and\
  \bibinfo {author} {\bibfnamefont {J.}~\bibnamefont {Berges}},\ }\bibfield
  {title} {\bibinfo {title} {Schwinger pair production with ultracold atoms},\
  }\href {https://doi.org/10.1016/j.physletb.2016.07.036} {\bibfield  {journal}
  {\bibinfo  {journal} {Physics Letters B}\ }\textbf {\bibinfo {volume}
  {760}},\ \bibinfo {pages} {742} (\bibinfo {year} {2016})}\BibitemShut
  {NoStop}%
\bibitem [{\citenamefont {Hu}\ \emph {et~al.}(2019)\citenamefont {Hu},
  \citenamefont {Feng}, \citenamefont {Zhang},\ and\ \citenamefont
  {Chin}}]{Hu_NatPhys_2019}%
  \BibitemOpen
  \bibfield  {author} {\bibinfo {author} {\bibfnamefont {J.}~\bibnamefont
  {Hu}}, \bibinfo {author} {\bibfnamefont {L.}~\bibnamefont {Feng}}, \bibinfo
  {author} {\bibfnamefont {Z.}~\bibnamefont {Zhang}},\ and\ \bibinfo {author}
  {\bibfnamefont {C.}~\bibnamefont {Chin}},\ }\bibfield  {title} {\bibinfo
  {title} {Quantum simulation of unruh radiation},\ }\href
  {https://doi.org/10.1038/s41567-019-0537-1} {\bibfield  {journal} {\bibinfo
  {journal} {Nature Physics}\ }\textbf {\bibinfo {volume} {15}},\ \bibinfo
  {pages} {785} (\bibinfo {year} {2019})}\BibitemShut {NoStop}%
\bibitem [{\citenamefont {Gross}\ \emph {et~al.}(2011)\citenamefont {Gross},
  \citenamefont {Strobel}, \citenamefont {Nicklas}, \citenamefont {Zibold},
  \citenamefont {Bar-Gill}, \citenamefont {Kurizki},\ and\ \citenamefont
  {Oberthaler}}]{Gross_Nature_2011}%
  \BibitemOpen
  \bibfield  {author} {\bibinfo {author} {\bibfnamefont {C.}~\bibnamefont
  {Gross}}, \bibinfo {author} {\bibfnamefont {H.}~\bibnamefont {Strobel}},
  \bibinfo {author} {\bibfnamefont {E.}~\bibnamefont {Nicklas}}, \bibinfo
  {author} {\bibfnamefont {T.}~\bibnamefont {Zibold}}, \bibinfo {author}
  {\bibfnamefont {N.}~\bibnamefont {Bar-Gill}}, \bibinfo {author}
  {\bibfnamefont {G.}~\bibnamefont {Kurizki}},\ and\ \bibinfo {author}
  {\bibfnamefont {M.~K.}\ \bibnamefont {Oberthaler}},\ }\bibfield  {title}
  {\bibinfo {title} {Atomic homodyne detection of continuous-variable entangled
  twin-atom states},\ }\href {https://doi.org/10.1038/nature10654} {\bibfield
  {journal} {\bibinfo  {journal} {Nature}\ }\textbf {\bibinfo {volume} {480}},\
  \bibinfo {pages} {219} (\bibinfo {year} {2011})}\BibitemShut {NoStop}%
\bibitem [{\citenamefont {L\"{u}cke}\ \emph {et~al.}(2011)\citenamefont
  {L\"{u}cke}, \citenamefont {Scherer}, \citenamefont {Kruse}, \citenamefont
  {Pezz{\'{e}}}, \citenamefont {Deuretzbacher}, \citenamefont {Hyllus},
  \citenamefont {Topic}, \citenamefont {Peise}, \citenamefont {Ertmer},
  \citenamefont {Arlt}, \citenamefont {Santos}, \citenamefont {Smerzi},\ and\
  \citenamefont {Klempt}}]{Lcke_Science_2011}%
  \BibitemOpen
  \bibfield  {author} {\bibinfo {author} {\bibfnamefont {B.}~\bibnamefont
  {L\"{u}cke}}, \bibinfo {author} {\bibfnamefont {M.}~\bibnamefont {Scherer}},
  \bibinfo {author} {\bibfnamefont {J.}~\bibnamefont {Kruse}}, \bibinfo
  {author} {\bibfnamefont {L.}~\bibnamefont {Pezz{\'{e}}}}, \bibinfo {author}
  {\bibfnamefont {F.}~\bibnamefont {Deuretzbacher}}, \bibinfo {author}
  {\bibfnamefont {P.}~\bibnamefont {Hyllus}}, \bibinfo {author} {\bibfnamefont
  {O.}~\bibnamefont {Topic}}, \bibinfo {author} {\bibfnamefont
  {J.}~\bibnamefont {Peise}}, \bibinfo {author} {\bibfnamefont
  {W.}~\bibnamefont {Ertmer}}, \bibinfo {author} {\bibfnamefont
  {J.}~\bibnamefont {Arlt}}, \bibinfo {author} {\bibfnamefont {L.}~\bibnamefont
  {Santos}}, \bibinfo {author} {\bibfnamefont {A.}~\bibnamefont {Smerzi}},\
  and\ \bibinfo {author} {\bibfnamefont {C.}~\bibnamefont {Klempt}},\
  }\bibfield  {title} {\bibinfo {title} {Twin matter waves for interferometry
  beyond the classical limit},\ }\href
  {https://doi.org/10.1126/science.1208798} {\bibfield  {journal} {\bibinfo
  {journal} {Science}\ }\textbf {\bibinfo {volume} {334}},\ \bibinfo {pages}
  {773} (\bibinfo {year} {2011})}\BibitemShut {NoStop}%
\bibitem [{\citenamefont {Linnemann}\ \emph {et~al.}(2016)\citenamefont
  {Linnemann}, \citenamefont {Strobel}, \citenamefont {Muessel}, \citenamefont
  {Schulz}, \citenamefont {Lewis-Swan}, \citenamefont {Kheruntsyan},\ and\
  \citenamefont {Oberthaler}}]{Linneman_PRL_2016}%
  \BibitemOpen
  \bibfield  {author} {\bibinfo {author} {\bibfnamefont {D.}~\bibnamefont
  {Linnemann}}, \bibinfo {author} {\bibfnamefont {H.}~\bibnamefont {Strobel}},
  \bibinfo {author} {\bibfnamefont {W.}~\bibnamefont {Muessel}}, \bibinfo
  {author} {\bibfnamefont {J.}~\bibnamefont {Schulz}}, \bibinfo {author}
  {\bibfnamefont {R.~J.}\ \bibnamefont {Lewis-Swan}}, \bibinfo {author}
  {\bibfnamefont {K.~V.}\ \bibnamefont {Kheruntsyan}},\ and\ \bibinfo {author}
  {\bibfnamefont {M.~K.}\ \bibnamefont {Oberthaler}},\ }\bibfield  {title}
  {\bibinfo {title} {Quantum-enhanced sensing based on time reversal of
  nonlinear dynamics},\ }\href {https://doi.org/10.1103/PhysRevLett.117.013001}
  {\bibfield  {journal} {\bibinfo  {journal} {Phys. Rev. Lett.}\ }\textbf
  {\bibinfo {volume} {117}},\ \bibinfo {pages} {013001} (\bibinfo {year}
  {2016})}\BibitemShut {NoStop}%
\bibitem [{\citenamefont {Qu}\ \emph {et~al.}(2020)\citenamefont {Qu},
  \citenamefont {Evrard}, \citenamefont {Dalibard},\ and\ \citenamefont
  {Gerbier}}]{Qu_PRL_2020}%
  \BibitemOpen
  \bibfield  {author} {\bibinfo {author} {\bibfnamefont {A.}~\bibnamefont
  {Qu}}, \bibinfo {author} {\bibfnamefont {B.}~\bibnamefont {Evrard}}, \bibinfo
  {author} {\bibfnamefont {J.}~\bibnamefont {Dalibard}},\ and\ \bibinfo
  {author} {\bibfnamefont {F.}~\bibnamefont {Gerbier}},\ }\bibfield  {title}
  {\bibinfo {title} {Probing spin correlations in a bose-einstein condensate
  near the single-atom level},\ }\href
  {https://doi.org/10.1103/PhysRevLett.125.033401} {\bibfield  {journal}
  {\bibinfo  {journal} {Phys. Rev. Lett.}\ }\textbf {\bibinfo {volume} {125}},\
  \bibinfo {pages} {033401} (\bibinfo {year} {2020})}\BibitemShut {NoStop}%
\bibitem [{\citenamefont {Chapman}\ \emph {et~al.}(2019)\citenamefont
  {Chapman}, \citenamefont {Eisert}, \citenamefont {Hackl}, \citenamefont
  {Heller}, \citenamefont {Jefferson}, \citenamefont {Marrochio},\ and\
  \citenamefont {Myers}}]{Chapman_SciPost_2019}%
  \BibitemOpen
  \bibfield  {author} {\bibinfo {author} {\bibfnamefont {S.}~\bibnamefont
  {Chapman}}, \bibinfo {author} {\bibfnamefont {J.}~\bibnamefont {Eisert}},
  \bibinfo {author} {\bibfnamefont {L.}~\bibnamefont {Hackl}}, \bibinfo
  {author} {\bibfnamefont {M.~P.}\ \bibnamefont {Heller}}, \bibinfo {author}
  {\bibfnamefont {R.}~\bibnamefont {Jefferson}}, \bibinfo {author}
  {\bibfnamefont {H.}~\bibnamefont {Marrochio}},\ and\ \bibinfo {author}
  {\bibfnamefont {R.~C.}\ \bibnamefont {Myers}},\ }\bibfield  {title} {\bibinfo
  {title} {Complexity and entanglement for thermofield double states},\
  }\bibfield  {journal} {\bibinfo  {journal} {{SciPost} Physics}\ }\textbf
  {\bibinfo {volume} {6}},\ \href
  {https://doi.org/10.21468/scipostphys.6.3.034} {10.21468/scipostphys.6.3.034}
  (\bibinfo {year} {2019})\BibitemShut {NoStop}%
\bibitem [{\citenamefont {Zhu}\ \emph {et~al.}(2020)\citenamefont {Zhu},
  \citenamefont {Johri}, \citenamefont {Linke}, \citenamefont {Landsman},
  \citenamefont {Alderete}, \citenamefont {Nguyen}, \citenamefont {Matsuura},
  \citenamefont {Hsieh},\ and\ \citenamefont {Monroe}}]{Zhu_PNAS_2020}%
  \BibitemOpen
  \bibfield  {author} {\bibinfo {author} {\bibfnamefont {D.}~\bibnamefont
  {Zhu}}, \bibinfo {author} {\bibfnamefont {S.}~\bibnamefont {Johri}}, \bibinfo
  {author} {\bibfnamefont {N.~M.}\ \bibnamefont {Linke}}, \bibinfo {author}
  {\bibfnamefont {K.~A.}\ \bibnamefont {Landsman}}, \bibinfo {author}
  {\bibfnamefont {C.~H.}\ \bibnamefont {Alderete}}, \bibinfo {author}
  {\bibfnamefont {N.~H.}\ \bibnamefont {Nguyen}}, \bibinfo {author}
  {\bibfnamefont {A.~Y.}\ \bibnamefont {Matsuura}}, \bibinfo {author}
  {\bibfnamefont {T.~H.}\ \bibnamefont {Hsieh}},\ and\ \bibinfo {author}
  {\bibfnamefont {C.}~\bibnamefont {Monroe}},\ }\bibfield  {title} {\bibinfo
  {title} {Generation of thermofield double states and critical ground states
  with a quantum computer},\ }\href {https://doi.org/10.1073/pnas.2006337117}
  {\bibfield  {journal} {\bibinfo  {journal} {Proceedings of the National
  Academy of Sciences}\ }\textbf {\bibinfo {volume} {117}},\ \bibinfo {pages}
  {25402} (\bibinfo {year} {2020})}\BibitemShut {NoStop}%
\bibitem [{\citenamefont {Tobias}\ \emph {et~al.}(2022)\citenamefont {Tobias},
  \citenamefont {Matsuda}, \citenamefont {Li}, \citenamefont {Miller},
  \citenamefont {Carroll}, \citenamefont {Bilitewski}, \citenamefont {Rey},\
  and\ \citenamefont {Ye}}]{Tobias_Science_375_2022}%
  \BibitemOpen
  \bibfield  {author} {\bibinfo {author} {\bibfnamefont {W.~G.}\ \bibnamefont
  {Tobias}}, \bibinfo {author} {\bibfnamefont {K.}~\bibnamefont {Matsuda}},
  \bibinfo {author} {\bibfnamefont {J.-R.}\ \bibnamefont {Li}}, \bibinfo
  {author} {\bibfnamefont {C.}~\bibnamefont {Miller}}, \bibinfo {author}
  {\bibfnamefont {A.~N.}\ \bibnamefont {Carroll}}, \bibinfo {author}
  {\bibfnamefont {T.}~\bibnamefont {Bilitewski}}, \bibinfo {author}
  {\bibfnamefont {A.~M.}\ \bibnamefont {Rey}},\ and\ \bibinfo {author}
  {\bibfnamefont {J.}~\bibnamefont {Ye}},\ }\bibfield  {title} {\bibinfo
  {title} {Reactions between layer-resolved molecules mediated by dipolar spin
  exchange},\ }\href {https://doi.org/10.1126/science.abn8525} {\bibfield
  {journal} {\bibinfo  {journal} {Science}\ }\textbf {\bibinfo {volume}
  {375}},\ \bibinfo {pages} {1299} (\bibinfo {year} {2022})}\BibitemShut
  {NoStop}%
\bibitem [{\citenamefont {Hammerer}\ \emph {et~al.}(2010)\citenamefont
  {Hammerer}, \citenamefont {S\o{}rensen},\ and\ \citenamefont
  {Polzik}}]{Polzik_RevModPhys_2010}%
  \BibitemOpen
  \bibfield  {author} {\bibinfo {author} {\bibfnamefont {K.}~\bibnamefont
  {Hammerer}}, \bibinfo {author} {\bibfnamefont {A.~S.}\ \bibnamefont
  {S\o{}rensen}},\ and\ \bibinfo {author} {\bibfnamefont {E.~S.}\ \bibnamefont
  {Polzik}},\ }\bibfield  {title} {\bibinfo {title} {Quantum interface between
  light and atomic ensembles},\ }\href
  {https://doi.org/10.1103/RevModPhys.82.1041} {\bibfield  {journal} {\bibinfo
  {journal} {Rev. Mod. Phys.}\ }\textbf {\bibinfo {volume} {82}},\ \bibinfo
  {pages} {1041} (\bibinfo {year} {2010})}\BibitemShut {NoStop}%
\bibitem [{\citenamefont {Freedman}\ and\ \citenamefont
  {Clauser}(1972)}]{freedman1972experimental}%
  \BibitemOpen
  \bibfield  {author} {\bibinfo {author} {\bibfnamefont {S.~J.}\ \bibnamefont
  {Freedman}}\ and\ \bibinfo {author} {\bibfnamefont {J.~F.}\ \bibnamefont
  {Clauser}},\ }\bibfield  {title} {\bibinfo {title} {Experimental test of
  local hidden-variable theories},\ }\href
  {https://doi.org/10.1103/PhysRevLett.28.938} {\bibfield  {journal} {\bibinfo
  {journal} {Phys. Rev. Lett.}\ }\textbf {\bibinfo {volume} {28}},\ \bibinfo
  {pages} {938} (\bibinfo {year} {1972})}\BibitemShut {NoStop}%
\bibitem [{\citenamefont {Aspect}\ \emph {et~al.}(1981)\citenamefont {Aspect},
  \citenamefont {Grangier},\ and\ \citenamefont
  {Roger}}]{aspect1981experimental}%
  \BibitemOpen
  \bibfield  {author} {\bibinfo {author} {\bibfnamefont {A.}~\bibnamefont
  {Aspect}}, \bibinfo {author} {\bibfnamefont {P.}~\bibnamefont {Grangier}},\
  and\ \bibinfo {author} {\bibfnamefont {G.}~\bibnamefont {Roger}},\ }\bibfield
   {title} {\bibinfo {title} {Experimental tests of realistic local theories
  via bell's theorem},\ }\href {https://doi.org/10.1103/PhysRevLett.47.460}
  {\bibfield  {journal} {\bibinfo  {journal} {Phys. Rev. Lett.}\ }\textbf
  {\bibinfo {volume} {47}},\ \bibinfo {pages} {460} (\bibinfo {year}
  {1981})}\BibitemShut {NoStop}%
\bibitem [{\citenamefont {Ou}\ \emph {et~al.}(1992)\citenamefont {Ou},
  \citenamefont {Pereira}, \citenamefont {Kimble},\ and\ \citenamefont
  {Peng}}]{ou1992realization}%
  \BibitemOpen
  \bibfield  {author} {\bibinfo {author} {\bibfnamefont {Z.~Y.}\ \bibnamefont
  {Ou}}, \bibinfo {author} {\bibfnamefont {S.~F.}\ \bibnamefont {Pereira}},
  \bibinfo {author} {\bibfnamefont {H.~J.}\ \bibnamefont {Kimble}},\ and\
  \bibinfo {author} {\bibfnamefont {K.~C.}\ \bibnamefont {Peng}},\ }\bibfield
  {title} {\bibinfo {title} {Realization of the einstein-podolsky-rosen paradox
  for continuous variables},\ }\href
  {https://doi.org/10.1103/PhysRevLett.68.3663} {\bibfield  {journal} {\bibinfo
   {journal} {Phys. Rev. Lett.}\ }\textbf {\bibinfo {volume} {68}},\ \bibinfo
  {pages} {3663} (\bibinfo {year} {1992})}\BibitemShut {NoStop}%
\bibitem [{\citenamefont {Julsgaard}\ \emph {et~al.}(2001)\citenamefont
  {Julsgaard}, \citenamefont {Kozhekin},\ and\ \citenamefont
  {Polzik}}]{julsgaard2001experimental}%
  \BibitemOpen
  \bibfield  {author} {\bibinfo {author} {\bibfnamefont {B.}~\bibnamefont
  {Julsgaard}}, \bibinfo {author} {\bibfnamefont {A.}~\bibnamefont
  {Kozhekin}},\ and\ \bibinfo {author} {\bibfnamefont {E.~S.}\ \bibnamefont
  {Polzik}},\ }\bibfield  {title} {\bibinfo {title} {Experimental long-lived
  entanglement of two macroscopic objects},\ }\href
  {https://doi.org/10.1038/35096524} {\bibfield  {journal} {\bibinfo  {journal}
  {Nature}\ }\textbf {\bibinfo {volume} {413}},\ \bibinfo {pages} {400}
  (\bibinfo {year} {2001})}\BibitemShut {NoStop}%
\bibitem [{\citenamefont {Cerf}\ \emph {et~al.}(2007)\citenamefont {Cerf},
  \citenamefont {Leuchs},\ and\ \citenamefont {Polzik}}]{cerf2007quantum}%
  \BibitemOpen
  \bibfield  {author} {\bibinfo {author} {\bibfnamefont {N.~J.}\ \bibnamefont
  {Cerf}}, \bibinfo {author} {\bibfnamefont {G.}~\bibnamefont {Leuchs}},\ and\
  \bibinfo {author} {\bibfnamefont {E.~S.}\ \bibnamefont {Polzik}},\ }\href
  {https://doi.org/10.1142/p489} {\emph {\bibinfo {title} {Quantum Information
  with Continuous Variables of Atoms and Light}}}\ (\bibinfo  {publisher}
  {World Scientific},\ \bibinfo {year} {2007})\BibitemShut {NoStop}%
\bibitem [{\citenamefont {Fadel}\ \emph {et~al.}(2017)\citenamefont {Fadel},
  \citenamefont {Zibold}, \citenamefont {Décamps},\ and\ \citenamefont
  {Treutlein}}]{Fadel2017}%
  \BibitemOpen
  \bibfield  {author} {\bibinfo {author} {\bibfnamefont {M.}~\bibnamefont
  {Fadel}}, \bibinfo {author} {\bibfnamefont {T.}~\bibnamefont {Zibold}},
  \bibinfo {author} {\bibfnamefont {B.}~\bibnamefont {Décamps}},\ and\
  \bibinfo {author} {\bibfnamefont {P.}~\bibnamefont {Treutlein}},\ }\bibfield
  {title} {\bibinfo {title} {Spatial entanglement patterns and
  einstein-podolsky-rosen steering in a bose-einstein condensate},\ }\href
  {https://doi.org/10.1126/science.aao1850} {\bibfield  {journal} {\bibinfo
  {journal} {Science}\ }\textbf {\bibinfo {volume} {360}} (\bibinfo {year}
  {2017})}\BibitemShut {NoStop}%
\bibitem [{\citenamefont {Kunkel}\ \emph {et~al.}(2018)\citenamefont {Kunkel},
  \citenamefont {Prüfer}, \citenamefont {Strobel}, \citenamefont {Linnemann},
  \citenamefont {Frölian}, \citenamefont {Gasenzer}, \citenamefont
  {Gärttner},\ and\ \citenamefont {Oberthaler}}]{Kunkel2018}%
  \BibitemOpen
  \bibfield  {author} {\bibinfo {author} {\bibfnamefont {P.}~\bibnamefont
  {Kunkel}}, \bibinfo {author} {\bibfnamefont {M.}~\bibnamefont {Prüfer}},
  \bibinfo {author} {\bibfnamefont {H.}~\bibnamefont {Strobel}}, \bibinfo
  {author} {\bibfnamefont {D.}~\bibnamefont {Linnemann}}, \bibinfo {author}
  {\bibfnamefont {A.}~\bibnamefont {Frölian}}, \bibinfo {author}
  {\bibfnamefont {T.}~\bibnamefont {Gasenzer}}, \bibinfo {author}
  {\bibfnamefont {M.}~\bibnamefont {Gärttner}},\ and\ \bibinfo {author}
  {\bibfnamefont {M.~K.}\ \bibnamefont {Oberthaler}},\ }\bibfield  {title}
  {\bibinfo {title} {Spatially distributed multipartite entanglement enables
  epr steering of atomic clouds},\ }\href
  {https://doi.org/10.1126/science.aao2254} {\bibfield  {journal} {\bibinfo
  {journal} {Science}\ }\textbf {\bibinfo {volume} {360}},\ \bibinfo {pages}
  {413} (\bibinfo {year} {2018})}\BibitemShut {NoStop}%
\bibitem [{\citenamefont {Stephenson}\ \emph {et~al.}(2020)\citenamefont
  {Stephenson}, \citenamefont {Nadlinger}, \citenamefont {Nichol},
  \citenamefont {An}, \citenamefont {Drmota}, \citenamefont {Ballance},
  \citenamefont {Thirumalai}, \citenamefont {Goodwin}, \citenamefont {Lucas},\
  and\ \citenamefont {Ballance}}]{PhysRevLett.124.110501}%
  \BibitemOpen
  \bibfield  {author} {\bibinfo {author} {\bibfnamefont {L.~J.}\ \bibnamefont
  {Stephenson}}, \bibinfo {author} {\bibfnamefont {D.~P.}\ \bibnamefont
  {Nadlinger}}, \bibinfo {author} {\bibfnamefont {B.~C.}\ \bibnamefont
  {Nichol}}, \bibinfo {author} {\bibfnamefont {S.}~\bibnamefont {An}}, \bibinfo
  {author} {\bibfnamefont {P.}~\bibnamefont {Drmota}}, \bibinfo {author}
  {\bibfnamefont {T.~G.}\ \bibnamefont {Ballance}}, \bibinfo {author}
  {\bibfnamefont {K.}~\bibnamefont {Thirumalai}}, \bibinfo {author}
  {\bibfnamefont {J.~F.}\ \bibnamefont {Goodwin}}, \bibinfo {author}
  {\bibfnamefont {D.~M.}\ \bibnamefont {Lucas}},\ and\ \bibinfo {author}
  {\bibfnamefont {C.~J.}\ \bibnamefont {Ballance}},\ }\bibfield  {title}
  {\bibinfo {title} {High-rate, high-fidelity entanglement of qubits across an
  elementary quantum network},\ }\href
  {https://doi.org/10.1103/PhysRevLett.124.110501} {\bibfield  {journal}
  {\bibinfo  {journal} {Phys. Rev. Lett.}\ }\textbf {\bibinfo {volume} {124}},\
  \bibinfo {pages} {110501} (\bibinfo {year} {2020})}\BibitemShut {NoStop}%
\bibitem [{\citenamefont {Nichol}\ \emph {et~al.}(2022)\citenamefont {Nichol},
  \citenamefont {Srinivas}, \citenamefont {Nadlinger}, \citenamefont {Drmota},
  \citenamefont {Main}, \citenamefont {Araneda}, \citenamefont {Ballance},\
  and\ \citenamefont {Lucas}}]{Nichol_2022}%
  \BibitemOpen
  \bibfield  {author} {\bibinfo {author} {\bibfnamefont {B.~C.}\ \bibnamefont
  {Nichol}}, \bibinfo {author} {\bibfnamefont {R.}~\bibnamefont {Srinivas}},
  \bibinfo {author} {\bibfnamefont {D.~P.}\ \bibnamefont {Nadlinger}}, \bibinfo
  {author} {\bibfnamefont {P.}~\bibnamefont {Drmota}}, \bibinfo {author}
  {\bibfnamefont {D.}~\bibnamefont {Main}}, \bibinfo {author} {\bibfnamefont
  {G.}~\bibnamefont {Araneda}}, \bibinfo {author} {\bibfnamefont {C.~J.}\
  \bibnamefont {Ballance}},\ and\ \bibinfo {author} {\bibfnamefont {D.~M.}\
  \bibnamefont {Lucas}},\ }\bibfield  {title} {\bibinfo {title} {An elementary
  quantum network of entangled optical atomic clocks},\ }\href
  {https://doi.org/10.1038/s41586-022-05088-z} {\bibfield  {journal} {\bibinfo
  {journal} {Nature}\ }\textbf {\bibinfo {volume} {609}},\ \bibinfo {pages}
  {689} (\bibinfo {year} {2022})}\BibitemShut {NoStop}%
\bibitem [{\citenamefont {Wan}\ \emph {et~al.}(2019)\citenamefont {Wan},
  \citenamefont {Kienzler}, \citenamefont {Erickson}, \citenamefont {Mayer},
  \citenamefont {Tan}, \citenamefont {Wu}, \citenamefont {Vasconcelos},
  \citenamefont {Glancy}, \citenamefont {Knill}, \citenamefont {Wineland},
  \citenamefont {Wilson},\ and\ \citenamefont {Leibfried}}]{Wan_2019}%
  \BibitemOpen
  \bibfield  {author} {\bibinfo {author} {\bibfnamefont {Y.}~\bibnamefont
  {Wan}}, \bibinfo {author} {\bibfnamefont {D.}~\bibnamefont {Kienzler}},
  \bibinfo {author} {\bibfnamefont {S.~D.}\ \bibnamefont {Erickson}}, \bibinfo
  {author} {\bibfnamefont {K.~H.}\ \bibnamefont {Mayer}}, \bibinfo {author}
  {\bibfnamefont {T.~R.}\ \bibnamefont {Tan}}, \bibinfo {author} {\bibfnamefont
  {J.~J.}\ \bibnamefont {Wu}}, \bibinfo {author} {\bibfnamefont {H.~M.}\
  \bibnamefont {Vasconcelos}}, \bibinfo {author} {\bibfnamefont
  {S.}~\bibnamefont {Glancy}}, \bibinfo {author} {\bibfnamefont
  {E.}~\bibnamefont {Knill}}, \bibinfo {author} {\bibfnamefont {D.~J.}\
  \bibnamefont {Wineland}}, \bibinfo {author} {\bibfnamefont {A.~C.}\
  \bibnamefont {Wilson}},\ and\ \bibinfo {author} {\bibfnamefont
  {D.}~\bibnamefont {Leibfried}},\ }\bibfield  {title} {\bibinfo {title}
  {Quantum gate teleportation between separated qubits in a trapped-ion
  processor},\ }\href {https://doi.org/10.1126/science.aaw9415} {\bibfield
  {journal} {\bibinfo  {journal} {Science}\ }\textbf {\bibinfo {volume}
  {364}},\ \bibinfo {pages} {875} (\bibinfo {year} {2019})}\BibitemShut
  {NoStop}%
\bibitem [{\citenamefont {Lago-Rivera}\ \emph {et~al.}(2021)\citenamefont
  {Lago-Rivera}, \citenamefont {Grandi}, \citenamefont {Rakonjac},
  \citenamefont {Seri},\ and\ \citenamefont
  {de~Riedmatten}}]{Lago_Rivera_2021}%
  \BibitemOpen
  \bibfield  {author} {\bibinfo {author} {\bibfnamefont {D.}~\bibnamefont
  {Lago-Rivera}}, \bibinfo {author} {\bibfnamefont {S.}~\bibnamefont {Grandi}},
  \bibinfo {author} {\bibfnamefont {J.~V.}\ \bibnamefont {Rakonjac}}, \bibinfo
  {author} {\bibfnamefont {A.}~\bibnamefont {Seri}},\ and\ \bibinfo {author}
  {\bibfnamefont {H.}~\bibnamefont {de~Riedmatten}},\ }\bibfield  {title}
  {\bibinfo {title} {Telecom-heralded entanglement between multimode
  solid-state quantum memories},\ }\href
  {https://doi.org/10.1038/s41586-021-03481-8} {\bibfield  {journal} {\bibinfo
  {journal} {Nature}\ }\textbf {\bibinfo {volume} {594}},\ \bibinfo {pages}
  {37} (\bibinfo {year} {2021})}\BibitemShut {NoStop}%
\bibitem [{\citenamefont {Zhou}\ \emph {et~al.}(2011)\citenamefont {Zhou},
  \citenamefont {Porto},\ and\ \citenamefont {Sarma}}]{Zhou_2011}%
  \BibitemOpen
  \bibfield  {author} {\bibinfo {author} {\bibfnamefont {Q.}~\bibnamefont
  {Zhou}}, \bibinfo {author} {\bibfnamefont {J.~V.}\ \bibnamefont {Porto}},\
  and\ \bibinfo {author} {\bibfnamefont {S.~D.}\ \bibnamefont {Sarma}},\
  }\bibfield  {title} {\bibinfo {title} {Condensates induced by interband
  coupling in a double-well lattice},\ }\href
  {https://doi.org/10.1103/physrevb.83.195106} {\bibfield  {journal} {\bibinfo
  {journal} {Physical Review B}\ }\textbf {\bibinfo {volume} {83}},\ \bibinfo
  {pages} {19516} (\bibinfo {year} {2011})}\BibitemShut {NoStop}%
\bibitem [{\citenamefont {McDonald}\ \emph {et~al.}(2018)\citenamefont
  {McDonald}, \citenamefont {Pereg-Barnea},\ and\ \citenamefont
  {Clerk}}]{McDonald_PhysicalReviewX_8_2018}%
  \BibitemOpen
  \bibfield  {author} {\bibinfo {author} {\bibfnamefont {A.}~\bibnamefont
  {McDonald}}, \bibinfo {author} {\bibfnamefont {T.}~\bibnamefont
  {Pereg-Barnea}},\ and\ \bibinfo {author} {\bibfnamefont {A.~A.}\ \bibnamefont
  {Clerk}},\ }\bibfield  {title} {\bibinfo {title} {Phase-dependent chiral
  transport and effective non-hermitian dynamics in a bosonic kitaev-majorana
  chain},\ }\href {https://doi.org/10.1103/physrevx.8.041031} {\bibfield
  {journal} {\bibinfo  {journal} {Physical Review X}\ }\textbf {\bibinfo
  {volume} {8}},\ \bibinfo {pages} {041031} (\bibinfo {year}
  {2018})}\BibitemShut {NoStop}%
\bibitem [{\citenamefont {Kitaev}(2001)}]{Kitaev_2001}%
  \BibitemOpen
  \bibfield  {author} {\bibinfo {author} {\bibfnamefont {A.~Y.}\ \bibnamefont
  {Kitaev}},\ }\bibfield  {title} {\bibinfo {title} {Unpaired majorana fermions
  in quantum wires},\ }\href {https://doi.org/10.1070/1063-7869/44/10S/S29}
  {\bibfield  {journal} {\bibinfo  {journal} {Physics-Uspekhi}\ }\textbf
  {\bibinfo {volume} {44}},\ \bibinfo {pages} {131} (\bibinfo {year}
  {2001})}\BibitemShut {NoStop}%
\bibitem [{\citenamefont {Perlin}\ \emph {et~al.}(2020)\citenamefont {Perlin},
  \citenamefont {Qu},\ and\ \citenamefont
  {Rey}}]{Perlin_PhysicalReviewLetters_125_2020}%
  \BibitemOpen
  \bibfield  {author} {\bibinfo {author} {\bibfnamefont {M.~A.}\ \bibnamefont
  {Perlin}}, \bibinfo {author} {\bibfnamefont {C.}~\bibnamefont {Qu}},\ and\
  \bibinfo {author} {\bibfnamefont {A.~M.}\ \bibnamefont {Rey}},\ }\bibfield
  {title} {\bibinfo {title} {Spin squeezing with short-range spin-exchange
  interactions},\ }\href {https://doi.org/10.1103/physrevlett.125.223401}
  {\bibfield  {journal} {\bibinfo  {journal} {Physical Review Letters}\
  }\textbf {\bibinfo {volume} {125}},\ \bibinfo {pages} {223401} (\bibinfo
  {year} {2020})}\BibitemShut {NoStop}%
\bibitem [{\citenamefont {Rey}\ \emph {et~al.}(2008)\citenamefont {Rey},
  \citenamefont {Jiang}, \citenamefont {Fleischhauer}, \citenamefont {Demler},\
  and\ \citenamefont {Lukin}}]{Rey_Phys.Rev.A_77_2008}%
  \BibitemOpen
  \bibfield  {author} {\bibinfo {author} {\bibfnamefont {A.~M.}\ \bibnamefont
  {Rey}}, \bibinfo {author} {\bibfnamefont {L.}~\bibnamefont {Jiang}}, \bibinfo
  {author} {\bibfnamefont {M.}~\bibnamefont {Fleischhauer}}, \bibinfo {author}
  {\bibfnamefont {E.}~\bibnamefont {Demler}},\ and\ \bibinfo {author}
  {\bibfnamefont {M.~D.}\ \bibnamefont {Lukin}},\ }\bibfield  {title} {\bibinfo
  {title} {Many-body protected entanglement generation in interacting spin
  systems},\ }\href {https://doi.org/10.1103/PhysRevA.77.052305} {\bibfield
  {journal} {\bibinfo  {journal} {Phys. Rev. A}\ }\textbf {\bibinfo {volume}
  {77}},\ \bibinfo {pages} {052305} (\bibinfo {year} {2008})}\BibitemShut
  {NoStop}%
\bibitem [{\citenamefont {Cappellaro}\ and\ \citenamefont
  {Lukin}(2009)}]{Cappellaro2009}%
  \BibitemOpen
  \bibfield  {author} {\bibinfo {author} {\bibfnamefont {P.}~\bibnamefont
  {Cappellaro}}\ and\ \bibinfo {author} {\bibfnamefont {M.~D.}\ \bibnamefont
  {Lukin}},\ }\bibfield  {title} {\bibinfo {title} {Quantum correlation in
  disordered spin systems: Applications to magnetic sensing},\ }\href
  {https://doi.org/10.1103/physreva.80.032311} {\bibfield  {journal} {\bibinfo
  {journal} {Physical Review A}\ }\textbf {\bibinfo {volume} {80}},\ \bibinfo
  {pages} {032311} (\bibinfo {year} {2009})}\BibitemShut {NoStop}%
\bibitem [{\citenamefont {Kwasigroch}\ and\ \citenamefont
  {Cooper}(2014)}]{Kwasigroch2014}%
  \BibitemOpen
  \bibfield  {author} {\bibinfo {author} {\bibfnamefont {M.~P.}\ \bibnamefont
  {Kwasigroch}}\ and\ \bibinfo {author} {\bibfnamefont {N.~R.}\ \bibnamefont
  {Cooper}},\ }\bibfield  {title} {\bibinfo {title} {Bose-einstein condensation
  and many-body localization of rotational excitations of polar molecules
  following a microwave pulse},\ }\href
  {https://doi.org/10.1103/physreva.90.021605} {\bibfield  {journal} {\bibinfo
  {journal} {Physical Review A}\ }\textbf {\bibinfo {volume} {90}},\ \bibinfo
  {pages} {021605(R)} (\bibinfo {year} {2014})}\BibitemShut {NoStop}%
\bibitem [{\citenamefont {Davis}\ \emph {et~al.}(2020)\citenamefont {Davis},
  \citenamefont {Periwal}, \citenamefont {Cooper}, \citenamefont {Bentsen},
  \citenamefont {Evered}, \citenamefont {VanKirk},\ and\ \citenamefont
  {Schleier-Smith}}]{Davis2020}%
  \BibitemOpen
  \bibfield  {author} {\bibinfo {author} {\bibfnamefont {E.~J.}\ \bibnamefont
  {Davis}}, \bibinfo {author} {\bibfnamefont {A.}~\bibnamefont {Periwal}},
  \bibinfo {author} {\bibfnamefont {E.~S.}\ \bibnamefont {Cooper}}, \bibinfo
  {author} {\bibfnamefont {G.}~\bibnamefont {Bentsen}}, \bibinfo {author}
  {\bibfnamefont {S.~J.}\ \bibnamefont {Evered}}, \bibinfo {author}
  {\bibfnamefont {K.}~\bibnamefont {VanKirk}},\ and\ \bibinfo {author}
  {\bibfnamefont {M.~H.}\ \bibnamefont {Schleier-Smith}},\ }\bibfield  {title}
  {\bibinfo {title} {Protecting spin coherence in a tunable heisenberg model},\
  }\href {https://doi.org/10.1103/physrevlett.125.060402} {\bibfield  {journal}
  {\bibinfo  {journal} {Physical Review Letters}\ }\textbf {\bibinfo {volume}
  {125}},\ \bibinfo {pages} {060402} (\bibinfo {year} {2020})}\BibitemShut
  {NoStop}%
\bibitem [{\citenamefont {Schachenmayer}\ \emph {et~al.}(2015)\citenamefont
  {Schachenmayer}, \citenamefont {Pikovski},\ and\ \citenamefont
  {Rey}}]{Schachenmayer_Phys.Rev.X_5_2015}%
  \BibitemOpen
  \bibfield  {author} {\bibinfo {author} {\bibfnamefont {J.}~\bibnamefont
  {Schachenmayer}}, \bibinfo {author} {\bibfnamefont {A.}~\bibnamefont
  {Pikovski}},\ and\ \bibinfo {author} {\bibfnamefont {A.~M.}\ \bibnamefont
  {Rey}},\ }\bibfield  {title} {\bibinfo {title} {Many-body quantum spin
  dynamics with monte carlo trajectories on a discrete phase space},\ }\href
  {https://doi.org/10.1103/PhysRevX.5.011022} {\bibfield  {journal} {\bibinfo
  {journal} {Phys. Rev. X}\ }\textbf {\bibinfo {volume} {5}},\ \bibinfo {pages}
  {011022} (\bibinfo {year} {2015})}\BibitemShut {NoStop}%
\bibitem [{\citenamefont {Zhu}\ \emph {et~al.}(2019)\citenamefont {Zhu},
  \citenamefont {Rey},\ and\ \citenamefont
  {Schachenmayer}}]{Zhu_NewJournalofPhysics_21_2019}%
  \BibitemOpen
  \bibfield  {author} {\bibinfo {author} {\bibfnamefont {B.}~\bibnamefont
  {Zhu}}, \bibinfo {author} {\bibfnamefont {A.~M.}\ \bibnamefont {Rey}},\ and\
  \bibinfo {author} {\bibfnamefont {J.}~\bibnamefont {Schachenmayer}},\
  }\bibfield  {title} {\bibinfo {title} {A generalized phase space approach for
  solving quantum spin dynamics},\ }\href
  {https://doi.org/10.1088/1367-2630/ab354d} {\bibfield  {journal} {\bibinfo
  {journal} {New Journal of Physics}\ }\textbf {\bibinfo {volume} {21}},\
  \bibinfo {pages} {082001} (\bibinfo {year} {2019})}\BibitemShut {NoStop}%
\bibitem [{\citenamefont {Sundar}\ \emph {et~al.}(2019)\citenamefont {Sundar},
  \citenamefont {Wang},\ and\ \citenamefont {Hazzard}}]{Sundar2019}%
  \BibitemOpen
  \bibfield  {author} {\bibinfo {author} {\bibfnamefont {B.}~\bibnamefont
  {Sundar}}, \bibinfo {author} {\bibfnamefont {K.~C.}\ \bibnamefont {Wang}},\
  and\ \bibinfo {author} {\bibfnamefont {K.~R.~A.}\ \bibnamefont {Hazzard}},\
  }\bibfield  {title} {\bibinfo {title} {Analysis of continuous and discrete
  wigner approximations for spin dynamics},\ }\href
  {https://doi.org/10.1103/physreva.99.043627} {\bibfield  {journal} {\bibinfo
  {journal} {Physical Review A}\ }\textbf {\bibinfo {volume} {99}},\ \bibinfo
  {pages} {043627} (\bibinfo {year} {2019})}\BibitemShut {NoStop}%
\bibitem [{\citenamefont {Wurtz}\ \emph {et~al.}(2018)\citenamefont {Wurtz},
  \citenamefont {Polkovnikov},\ and\ \citenamefont {Sels}}]{Wurtz2018}%
  \BibitemOpen
  \bibfield  {author} {\bibinfo {author} {\bibfnamefont {J.}~\bibnamefont
  {Wurtz}}, \bibinfo {author} {\bibfnamefont {A.}~\bibnamefont {Polkovnikov}},\
  and\ \bibinfo {author} {\bibfnamefont {D.}~\bibnamefont {Sels}},\ }\bibfield
  {title} {\bibinfo {title} {Cluster truncated wigner approximation in strongly
  interacting systems},\ }\href
  {https://doi.org/https://doi.org/10.1016/j.aop.2018.06.001} {\bibfield
  {journal} {\bibinfo  {journal} {Annals of Physics}\ }\textbf {\bibinfo
  {volume} {395}},\ \bibinfo {pages} {341} (\bibinfo {year}
  {2018})}\BibitemShut {NoStop}%
\bibitem [{\citenamefont {Kolodrubetz}\ \emph {et~al.}(2017)\citenamefont
  {Kolodrubetz}, \citenamefont {Sels}, \citenamefont {Mehta},\ and\
  \citenamefont {Polkovnikov}}]{Kolodrubetz2017}%
  \BibitemOpen
  \bibfield  {author} {\bibinfo {author} {\bibfnamefont {M.}~\bibnamefont
  {Kolodrubetz}}, \bibinfo {author} {\bibfnamefont {D.}~\bibnamefont {Sels}},
  \bibinfo {author} {\bibfnamefont {P.}~\bibnamefont {Mehta}},\ and\ \bibinfo
  {author} {\bibfnamefont {A.}~\bibnamefont {Polkovnikov}},\ }\bibfield
  {title} {\bibinfo {title} {Geometry and non-adiabatic response in quantum and
  classical systems},\ }\href {https://doi.org/10.1016/j.physrep.2017.07.001}
  {\bibfield  {journal} {\bibinfo  {journal} {Physics Reports}\ }\textbf
  {\bibinfo {volume} {697}},\ \bibinfo {pages} {1} (\bibinfo {year}
  {2017})}\BibitemShut {NoStop}%
\bibitem [{\citenamefont {Berg}\ \emph {et~al.}(2009)\citenamefont {Berg},
  \citenamefont {Plimak}, \citenamefont {Polkovnikov}, \citenamefont {Olsen},
  \citenamefont {Fleischhauer},\ and\ \citenamefont {Schleich}}]{Berg2009}%
  \BibitemOpen
  \bibfield  {author} {\bibinfo {author} {\bibfnamefont {B.}~\bibnamefont
  {Berg}}, \bibinfo {author} {\bibfnamefont {L.~I.}\ \bibnamefont {Plimak}},
  \bibinfo {author} {\bibfnamefont {A.}~\bibnamefont {Polkovnikov}}, \bibinfo
  {author} {\bibfnamefont {M.~K.}\ \bibnamefont {Olsen}}, \bibinfo {author}
  {\bibfnamefont {M.}~\bibnamefont {Fleischhauer}},\ and\ \bibinfo {author}
  {\bibfnamefont {W.~P.}\ \bibnamefont {Schleich}},\ }\bibfield  {title}
  {\bibinfo {title} {Commuting heisenberg operators as the quantum response
  problem: Time-normal averages in the truncated wigner representation},\
  }\href {https://doi.org/10.1103/physreva.80.033624} {\bibfield  {journal}
  {\bibinfo  {journal} {Physical Review A}\ }\textbf {\bibinfo {volume} {80}},\
  \bibinfo {pages} {033624} (\bibinfo {year} {2009})}\BibitemShut {NoStop}%
\bibitem [{\citenamefont {Holstein}\ and\ \citenamefont
  {Primakoff}(1940)}]{Holstein1940}%
  \BibitemOpen
  \bibfield  {author} {\bibinfo {author} {\bibfnamefont {T.}~\bibnamefont
  {Holstein}}\ and\ \bibinfo {author} {\bibfnamefont {H.}~\bibnamefont
  {Primakoff}},\ }\bibfield  {title} {\bibinfo {title} {Field dependence of the
  intrinsic domain magnetization of a ferromagnet},\ }\href
  {https://doi.org/10.1103/physrev.58.1098} {\bibfield  {journal} {\bibinfo
  {journal} {Physical Review}\ }\textbf {\bibinfo {volume} {58}},\ \bibinfo
  {pages} {1098} (\bibinfo {year} {1940})}\BibitemShut {NoStop}%
\bibitem [{\citenamefont {{Sundar}}\ \emph {et~al.}(2022)\citenamefont
  {{Sundar}}, \citenamefont {{Barberena}}, \citenamefont {{Pineiro Orioli}},
  \citenamefont {{Chu}}, \citenamefont {{Thompson}}, \citenamefont {{Rey}},\
  and\ \citenamefont {{Lewis-Swan}}}]{Sundar2022}%
  \BibitemOpen
  \bibfield  {author} {\bibinfo {author} {\bibfnamefont {B.}~\bibnamefont
  {{Sundar}}}, \bibinfo {author} {\bibfnamefont {D.}~\bibnamefont
  {{Barberena}}}, \bibinfo {author} {\bibfnamefont {A.}~\bibnamefont {{Pineiro
  Orioli}}}, \bibinfo {author} {\bibfnamefont {A.}~\bibnamefont {{Chu}}},
  \bibinfo {author} {\bibfnamefont {J.~K.}\ \bibnamefont {{Thompson}}},
  \bibinfo {author} {\bibfnamefont {A.~M.}\ \bibnamefont {{Rey}}},\ and\
  \bibinfo {author} {\bibfnamefont {R.~J.}\ \bibnamefont {{Lewis-Swan}}},\
  }\bibfield  {title} {\bibinfo {title} {{Bosonic pair production and squeezing
  for optical phase measurements in long-lived dipoles coupled to a cavity}},\
  }\href@noop {} {\bibfield  {journal} {\bibinfo  {journal} {arXiv e-prints}\
  ,\ \bibinfo {eid} {arXiv:2204.13090}} (\bibinfo {year} {2022})},\ \Eprint
  {https://arxiv.org/abs/2204.13090} {arXiv:2204.13090 [quant-ph]} \BibitemShut
  {NoStop}%
\bibitem [{sup()}]{supplemental}%
  \BibitemOpen
  \href@noop {} {}\bibinfo {note} {See Supplemental Material which also
  contains Ref. 74 at [URL will be inserted by publisher] for additional
  details on the Holstein-Primakoff transformation for spiral states, the exact
  solution of the bosonic Kitaev model, and the comparison of the analytical
  results of the bosonic Kitaev model with the dipolar spin
  dynamics}\BibitemShut {NoStop}%
\bibitem [{\citenamefont {Rodriguez-Nieva}\ \emph {et~al.}(2022)\citenamefont
  {Rodriguez-Nieva}, \citenamefont {Orioli},\ and\ \citenamefont
  {Marino}}]{RodriguezNieva2022}%
  \BibitemOpen
  \bibfield  {author} {\bibinfo {author} {\bibfnamefont {J.~F.}\ \bibnamefont
  {Rodriguez-Nieva}}, \bibinfo {author} {\bibfnamefont {A.~P.}\ \bibnamefont
  {Orioli}},\ and\ \bibinfo {author} {\bibfnamefont {J.}~\bibnamefont
  {Marino}},\ }\bibfield  {title} {\bibinfo {title} {Far-from-equilibrium
  universality in the two-dimensional heisenberg model},\ }\bibfield  {journal}
  {\bibinfo  {journal} {Proceedings of the National Academy of Sciences}\
  }\textbf {\bibinfo {volume} {119}},\ \href
  {https://doi.org/10.1073/pnas.2122599119} {10.1073/pnas.2122599119} (\bibinfo
  {year} {2022})\BibitemShut {NoStop}%
\bibitem [{\citenamefont {G\"{a}rttner}\ \emph {et~al.}(2017)\citenamefont
  {G\"{a}rttner}, \citenamefont {Bohnet}, \citenamefont {Safavi-Naini},
  \citenamefont {Wall}, \citenamefont {Bollinger},\ and\ \citenamefont
  {Rey}}]{Grttner_NatPhys_2017}%
  \BibitemOpen
  \bibfield  {author} {\bibinfo {author} {\bibfnamefont {M.}~\bibnamefont
  {G\"{a}rttner}}, \bibinfo {author} {\bibfnamefont {J.~G.}\ \bibnamefont
  {Bohnet}}, \bibinfo {author} {\bibfnamefont {A.}~\bibnamefont
  {Safavi-Naini}}, \bibinfo {author} {\bibfnamefont {M.~L.}\ \bibnamefont
  {Wall}}, \bibinfo {author} {\bibfnamefont {J.~J.}\ \bibnamefont
  {Bollinger}},\ and\ \bibinfo {author} {\bibfnamefont {A.~M.}\ \bibnamefont
  {Rey}},\ }\bibfield  {title} {\bibinfo {title} {Measuring out-of-time-order
  correlations and multiple quantum spectra in a trapped-ion quantum magnet},\
  }\href {https://doi.org/10.1038/nphys4119} {\bibfield  {journal} {\bibinfo
  {journal} {Nature Physics}\ }\textbf {\bibinfo {volume} {13}},\ \bibinfo
  {pages} {781} (\bibinfo {year} {2017})}\BibitemShut {NoStop}%
\bibitem [{\citenamefont {dos Santos}\ \emph {et~al.}(2018)\citenamefont {dos
  Santos}, \citenamefont {dos Santos~Dias}, \citenamefont {Guimar{\~{a}}es},
  \citenamefont {Bouaziz},\ and\ \citenamefont {Lounis}}]{Santos2018}%
  \BibitemOpen
  \bibfield  {author} {\bibinfo {author} {\bibfnamefont {F.~J.}\ \bibnamefont
  {dos Santos}}, \bibinfo {author} {\bibfnamefont {M.}~\bibnamefont {dos
  Santos~Dias}}, \bibinfo {author} {\bibfnamefont {F.~S.~M.}\ \bibnamefont
  {Guimar{\~{a}}es}}, \bibinfo {author} {\bibfnamefont {J.}~\bibnamefont
  {Bouaziz}},\ and\ \bibinfo {author} {\bibfnamefont {S.}~\bibnamefont
  {Lounis}},\ }\bibfield  {title} {\bibinfo {title} {Spin-resolved inelastic
  electron scattering by spin waves in noncollinear magnets},\ }\href
  {https://doi.org/10.1103/PhysRevB.97.024431} {\bibfield  {journal} {\bibinfo
  {journal} {Phys. Rev. B}\ }\textbf {\bibinfo {volume} {97}},\ \bibinfo
  {pages} {024431} (\bibinfo {year} {2018})}\BibitemShut {NoStop}%
\end{thebibliography}


\cleardoublepage
\appendix
\section{Supplementary Information}

\subsection{Holstein-Primakoff for non-coplanar states}
In this section, we briefly recap the Holstein-Primakoff transformation applied to non-collinear states following Ref.\cite{Santos2018}. We start from a nearest-neighbor spin model 

\begin{equation*}
    H = 1/2 \sum_{ij} \sum_{\mu \nu} S_i^{\mu} J^{\mu \nu}_{ij} S_j^{\nu} = \sum_{i,j} S_i^T \mathbf{J}_{ij} S_j
\end{equation*}
where $i,j$ label sites/pancake indices, and $\mu,\nu$ label spin components $x,y,z$. 

We now perform site-dependent rotations to a local spin-coordinate frame 
\begin{equation}
    S_j = U^{\dagger}_{j} \tilde{S}_j U_j =\mathbf{R}_j S_j
\end{equation}
where $U_j = e^{ i \theta \left(\vec{S} \cdot \hat {n}\right)}$ is a general rotation and
 $\mathbf{R}_i$ is the corresponding rotation-matrix defined by the equality above. Thus, in the rotated frame we obtain

\begin{equation}
H = 1/2 \sum_{ij} \tilde{S}_i^T \mathbf{R}_i \mathbf{J}_{ij} \mathbf{R_j}^T \tilde{S}_j = 1/2 \sum_{ij}  \tilde{S}_i^T \tilde{\mathbf{J}}_{ij} \tilde{S}_j
\end{equation}
where we defined $\tilde{\mathbf{J}}_{ij}  =  \mathbf{R}_i \mathbf{J}_{ij} \mathbf{R_j}^T $.

Assuming that in the rotated spin basis the initial state is aligned along $Z$ on all sites $i$, $S_i(t=0) = (0,0,1)$, as a second step we now perform a Holstein-Primakoff transformation, 

\begin{equation}
    \tilde{S}_i = \mathbf{M} \mathbf{\hat{a}}_i
\end{equation}
with
\begin{equation}
    M = \sqrt{S/2} \begin{pmatrix}
                        1 & 1 & 0 \\
                        -i & i & 0\\
                        0 & 0 & \sqrt{2/S}
                 \end{pmatrix} , 
 \qquad  \hat{\mathbf{a}}_i = \begin{pmatrix}
                    \hat{a}_i\\
                    \hat{a}_i^{\dagger}\\
                    S - \hat{a}_i^{\dagger} a_i
                 \end{pmatrix}
\end{equation}

to obtain

\begin{equation*}
    H = 1/2 \sum_{ij}  \hat{\mathbf{a}}^{\dagger}_i \mathbf{J}^{'}_{ij} \hat{\mathbf{a}}_{j}
\end{equation*}

with
\begin{equation}
    \mathbf{J}_{ij}^{'} = \mathbf{M}^T \mathbf{R}_i \mathbf{J}_{ij} \mathbf{R}_j^T \mathbf{M}
\end{equation}

Now truncating the Hamiltonian at quadratic order in the bosonic operators we have

\begin{equation}
    H \approx H_0 + H_1 + H_2
\end{equation}

We ignore the zero-order term which corresponds to the classical energy of the initial state. 

The second order term can be written as 

\begin{align}
    H_2 = 1/2 \sum_{ij} \left(\hat{a}_i, \hat{a}^{\dagger}_i \right) \begin{pmatrix}
    H_{ij}^{++} & H_{ij}^{+-} \\
    H_{ij}^{-+} & H_{ij}^{--}
    \end{pmatrix}
    \begin{pmatrix}
    \hat{a}_j\\
    \hat{a}^{\dagger}_j
    \end{pmatrix}
\end{align}

The first order term takes the form 
\begin{equation}
    H_1 = \sum_i \begin{pmatrix}
                    h^+\\
                    h^-
                \end{pmatrix}
                \cdot
                \begin{pmatrix}
                    \hat{a}_i\\
                    \hat{a}^{\dagger}_i
                \end{pmatrix}
\end{equation}  

Generically, the first order term does not vanish for an arbitrary initial state that is not the classical groundstate of the model, and thus there will be mean-field dynamics.

\subsection{$(1,1,1)$ Spiral State}
\begin{figure}[t!]
\includegraphics[width=\columnwidth]{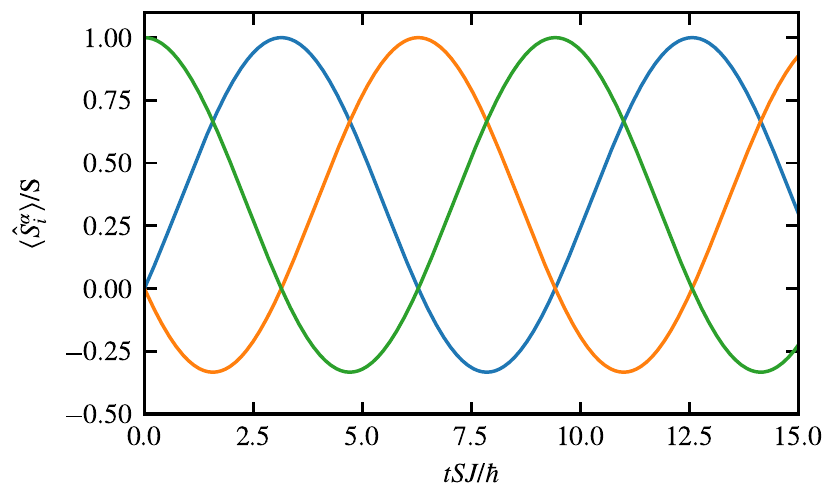}
\caption{Mean field dynamics of the $(1,1,1)$ spiral state. Shown are the $x,y,z$ components of the spins which are up to permutation the same for all spins (assuming that the lattice size is a multiple of 3 under periodic boundary conditions).}
\label{fig_sup:spiral_MF}
\end{figure} 
In the following we specialise to a translationally invariant system with periodic boundary conditions with nearest neighbour Heisenberg interactions, e.g. $\mathbf{J}^{\mu,\nu}_{i,j} =(\delta_{j,i+1} + \delta_{j,i-1}) \delta_{\mu,\nu} J$.
%

We consider a non-coplanar spiral, for which the initial state is polarised along $(x,y,z)$ on sites $1,2,3$ repeating periodically. This corresponds to $U_j = e^{ i j 2\pi /3 \left(\vec{S} \cdot \hat {n}\right)}$ with $\hat{n} = (1,1,1)/\sqrt{3}$ and $R_j$ being the rotation around $(1,1,1)/\sqrt{3}$ by an angle $j \, 2 \pi/3$.

The Hamiltonian in the rotated frame can be written as
\begin{equation}
    H = J \sum_i \left( S_i^x S_{i+1}^z + S_i^y S_{i+1}^x + S_{i}^z S_{i+1}^y \right)
\end{equation}

Substituting in the HP transformation, we see that in the rotated frame we find an effective magnetic field $h=(S J,S J,0)$, or alternatively $h^+ = S \sqrt{S/2}(J -i J) $, $h^-=S \sqrt{S/2} (J+iJ)$. 

Fortunately, this global rotation can easily be removed by a time-dependent unitary transformation. For simplicity we perform this rotation in the original frame via
\begin{equation}
    U(t) = e^{- i t J \sum_i \left(S_i^x +S_i^y +S_i^z\right)}
\end{equation}
which demonstrates that the mean field dynamics is just a global rotation of all spins around the $(1,1,1)$ direction.

We show in Fig.~\ref{fig_sup:spiral_MF} that the mean-field dynamics starting from this initial state indeed corresponds to a rigid rotation of all spins around the $(1,1,1)$ direction at a constant speed as seen in the collapse of all spin components of all sites along the chain onto only 3 curves which can for example be taken to be the $x,y,z$ components of a spin on any site.  Interestingly, this thus corresponds to a special initial state for which the dynamics of the (classical) Heisenberg chain can be exactly solved.

Since this unitary transformation commutes with the Hamiltonian (as it only depends on the total magnetisation which is preserved by the Heisenberg Hamiltonian), the Hamiltonian in the time-dependent rotating frame simplifies to

\begin{equation}
H_{U} = U H U^{\dagger} + i \hbar \frac{\partial U}{\partial t}U^{\dagger} = H - J \sum_{i} \left( S_i^x +S_i^y +S_i^z\right)
\label{eq_SM:H_U}
\end{equation}

Starting from this Hamiltonian in a time-dependent frame, performing the transformation to the initial state and the HP transformation we obtain the second order Hamiltonian as
\begin{align}
    H_U &\approx JS \sum_i \left(\hat{a}_i, \hat{a}^{\dagger}_i \right)\begin{pmatrix}
            i  & i \\
            -i  & -i
        \end{pmatrix}
        \begin{pmatrix}\hat{a}_{i+1} \\
        \hat{a}^{\dagger}_{i+1} 
        \end{pmatrix}\\
        &= JS \sum_i \left( i  \hat{a}_i \hat{a}^{\dagger}_{i+1}  - i \hat{a}^{\dagger}_i \hat{a}^{\dagger}_{i+1}  + \mathrm{H.c.} \right)
\end{align}
which is the "bosonic" kitaev model at the "critical" point. It turns out that the corresponding dynamical/associated matrix cannot be diagonalised as all eigenvalues are zero, but only 2 zero eigenvectors exist. 

It is more convenient to this model in terms of hermitian quadrature operators, $ \hat{a}_i = (\hat{x} + i \hat{p}_i)/\sqrt{2}$, $\hat{a}^{\dagger} = (\hat{x} -i \hat{p}_i)/\sqrt{2} $, which in this case up to rescaling are just the $\tilde{S}^x$ and $\tilde{S}^y$ components of the spin-vector in the time-dependent frame, resulting in

\begin{equation}
    H = -2 J S \sum_{i} \hat{p}_{i} \hat{x}_{i+1}
\end{equation}

From this we obtain the equations of motion for $\hat{c} = (\hat{x}_1,\cdots,\hat{x}_N,\hat{p}_1,\cdots,\hat{p_N})$ as $\partial_t \hat{c} = M \hat{c}$ with 
\begin{equation}
    M = -2 J S/\hbar \begin{pmatrix}
                      S  & 0\\
                  0  & -S^T
                \end{pmatrix}
\end{equation}
with 
\begin{equation}
S = 
\begin{pmatrix}
  & 0 & 1            &  0        & \cdots &0\\
  &   & \ddots       & \ddots    & \ddots & \vdots \\
  &   &              &\ddots     & \ddots & 0\\
  &   & \text{\huge0}&           & \ddots & 1 \\
  &   &              &           &        & 0
 \end{pmatrix}
\end{equation}
which only contains a single upper diagonal entry, e.g. is in Jordan normal form. This immediately allows us to compute the matrix exponential using standard formulae. We also note the explicit zero-modes given by $\hat{x}_N$ and $\hat{p}_1$ apparent in the zero row of S and immediate from the chiral coupling of the $\hat{x}$ and $\hat{p}$ operators and the open boundary conditions.

Defining $T =e^{t M} $ we have  $\hat{c}(t) = T  \hat{c}(t=0) $, and we can directly obtain the Green's function $ \left[ \hat{c}(t=0), \hat{c}^T(t) \right] = G T^T $
 with $G = \begin{pmatrix}
   \mathbf{0} & i \mathbf{1}\\
   -i \mathbf{1} & \mathbf{0}
 \end{pmatrix}$
 
 Based on this we obtain 

  \begin{align}
   \left[  \hat{x}_i, \hat{p}_{i+r}(t) \right] &=  \begin{cases}       i {(-2 J S t/\hbar)}^r/r!         & \text{if $ r \ge 0$}  \\
                                                                      0 & \text{if $ r < 0$} 
                                                    \end{cases}\label{eq_sup:greens_1}\\
      \left[  \hat{p}_i, \hat{x}_{i+r}(t) \right] &= \begin{cases}    0 & \text{if  $r > 0$} \\
                                                                          -i {(2 J S t /\hbar)}^r/r!       & \text{if  $r \le 0$}
                                                    \end{cases}\label{eq_sup:greens_2}\\
     \left[  \hat{x}_i, \hat{x}_{j}(t) \right] &=  \left[  \hat{p}_i, \hat{p}_{j}(t) \right] = 0                        
 \end{align}
These expressions are manifestly chiral, confirming the expectation that a forcing of $\hat{p}(t)$ due to $\hat{x}$ propagates only to the right (and vice-versa).
 
We note that the same calculation can of course be performed in the basis of bosonic operators, and our Hamiltonian falls into the class of non-diagonalisable bosonic problems due to soft modes. Specifically, all eigenvalues are 0, and there only are the two explicit zero-modes/eigenvectors mentioned, with $2N-2$ eigenvectors missing. 

One can also perform a short-time expansion of the full spin green's function
\begin{align}
    \left[\hat{S}^{\alpha}_i, \hat{S}^{\beta}_j(t) \right] &= \left[ \hat{S}^{\alpha}_i, e^{it H/\hbar} \hat{S}^{\beta}_j  e^{-it H/\hbar}\right] \\
    &\approx \left[\hat{S}^{\alpha}_i, \hat{S}^{\beta}_j \right] + \frac{it}{\hbar} \left[\hat{S}^{\alpha}_i , \left[ H, \hat{S}^{\beta}_j \right]\right]
\end{align}
which in the case of $i\neq j$ and $\alpha \neq \beta$ reduces to
\begin{equation}
     \left[\hat{S}^{\alpha}_i, \hat{S}^{\beta}_j(t) \right] \approx \frac{-i J t}{\hbar} V_{ij} \hat{S}^{\beta}_j \hat{S}^{\alpha}_i
\end{equation}
For the initial $(1,1,1)$ spiral state with sites pointing along $x,y,z$ periodically evaluating this on a site $i_z$ which points initially along $z$ we then obtain
\begin{align}
    \langle \left[\hat{S}_{i_z}^{x} , \hat{S}_{i_z+1}^z(t)\right] \rangle&\approx \frac{-i Jt}{\hbar} S^2 \\
   \langle \left[\hat{S}_{i_z}^{y} , \hat{S}_{i_z-1}^z(t)\right] \rangle &\approx \frac{-i Jt}{\hbar} S^2
\end{align}
which agrees with the results of Eq.~\ref{eq_sup:greens_1}-\ref{eq_sup:greens_2} in linear order after rotating to the proper spin-frame and using $\tilde{S}_x \approx \sqrt{S/2} \hat{x}$ and $\tilde{S}_Y \approx \sqrt{S/2} \hat{p}$. This provides a direct intuitive explanation of the chiral behavior in the spin language in terms of the local exchange fields exerted by nearest neighbours.
\subsection{Dipolar Multi-Layer - Spiral Dynamics}
\begin{figure}[!t]
\includegraphics[width=\columnwidth]{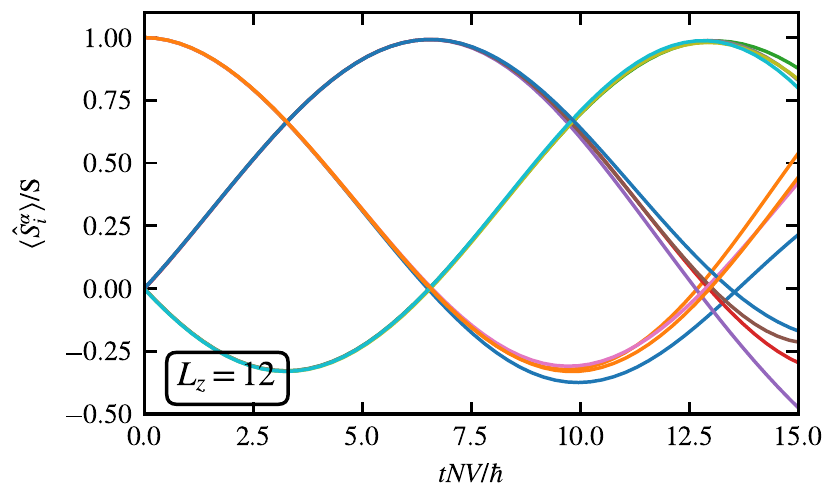}
\caption{dTWA dynamics of the dipolar multi-layer with $L_X=L_Y=30$, $L_Z=12$ system with $a_Z/a_{lat}=12$, initially prepared layer-selectively in the  $(1,1,1)$ spiral state. Shown are the $x,y,z$ components of the collective spin in each layer  $S^{\alpha}_i(t) = \sum_{i_X,i_Y} \langle\sigma_{i_X,i_Y,i}^{\alpha}(t)\rangle$ for layers $i=3,\cdots,9$. Note this assumes open boundary conditions and includes all the dipolar interactions}
\label{fig_sup:spiral_dTWA}
\end{figure} 
While the discussion so far strictly only applies in the case of the nearest neighbour Heisenberg spin model with periodic boundary conditions, corrections in the full model of dipoles prepared in multiple two-dimensional stacked layers with the full dipolar interactions remain relatively small. 

We show in Fig.~\ref{fig_sup:spiral_dTWA} the individual $x,y,z$ components of the collective layer spins $S^{\alpha}_i(t) = \sum_{i_X,i_Y} \sigma_{i_X,i_Y,i}^{\alpha}(t)$ of the full dipolar spin model including the full spatial dependence of the dipolar interactions and using open boundary conditions obtained via dTWA simulations. We note that due to the open boundary conditions and loss of translational invariance the individual layers are not equivalent up to permutation of the spin components anymore, which is the main reason for different layers showing distinct dynamics on these timescales. Even so the dynamics remains close to the ideal neareast neighbour case with periodic boundary conditions over a long time-scale. Deviation appear approximately at a time-scale of the order of $L_Z/2$ where perturbations are expected to have propagated from the edges to the central layer.

 \begin{figure}[!t]
\includegraphics[width=\columnwidth]{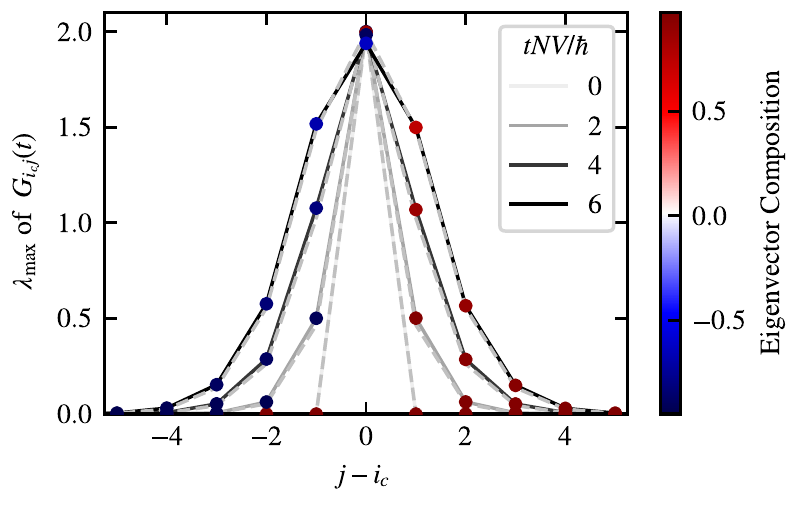}
\includegraphics[width=\columnwidth]{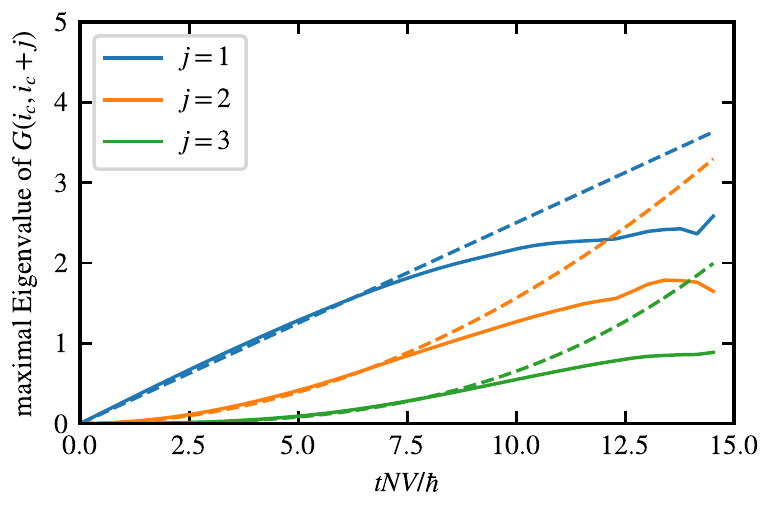}
\caption{Top: Largest eigenvalue $\lambda_{max}$ of the spin Green's function $G_{i_c,j}^{\alpha \beta}(t)=i\langle \left[S_{i_c}^{\alpha}(t=0), S_j^{\beta}(t) \right]\rangle$ at fixed times versus distance j. Colorbar shows the (left) eigenvector structure $|v^x|^2 -  |v^y|^2$ indicating that correlations that propagate to the right originate from $S_{i_c}^x(t=0)$, while correlations that propagate to the left originate from $S_{i_c}^y(t=0)$. Gray-dashed lines are the analytical results (Eqs.~\ref{eq_sup:greens_1}-\ref{eq_sup:greens_2}). Bottom: Largest eigenvalue of the spin Green's function $G^{\alpha \beta}_{i_c,j}(t)$ versus time for different distances $j$ (solid), compared to the analytical results (dashed) showing linear, quadratic and cubic dependence of $t$. All results for $L_X=L_Y=10$, $L_Z=12$ \label{fig_sup:spiral_kitaev}
}
\end{figure} 
 Next, we compare the analytical predictions based on the Holstein-Primakoff transformation of the ideal nearest-neighbour XXX spin chain from (Eqs.~\ref{eq_sup:greens_1}- \ref{eq_sup:greens_2}) to the full dynamics of the multi-layer system in Fig.~\ref{fig_sup:spiral_kitaev}. We observe both qualitative as well as decent quantitative agreement: the full dynamics shows the expected chirality (top panel), and the distance resolved results show the correct polynomial time-dependence (bottom panel) up to the times where boundary effects become important.

\end{document}